\documentclass[twocolumn,aps,prd,superscriptaddress,preprintnumbers,nofootinbib,10pt]{revtex4-2}
\usepackage{multirow}
\usepackage{amsmath}
\usepackage{amssymb}
\usepackage[dvipdf,dvips]{graphicx}
\usepackage{color}
\usepackage{hyperref}
\usepackage{url}
\usepackage{slashed}
\usepackage{subfigure}
\usepackage[usenames,dvipsnames]{xcolor}
\usepackage{amsmath}
\usepackage{amsfonts}
\usepackage{float} 
\usepackage{amssymb}
\usepackage{epsfig}
\usepackage{graphics}
\usepackage{euscript}
\usepackage{slashed}
\usepackage{epstopdf}
\usepackage[utf8]{inputenc}
\allowdisplaybreaks
\usepackage[normalem]{ulem}
\usepackage{pifont}
\usepackage{dsfont}
\usepackage{MnSymbol}
\usepackage{graphicx}
\usepackage{latexsym}
\usepackage{tikz-feynman}
\usepackage{easyReview}
\usepackage{soul}

\usepackage{dsfont}

\begin{document}
	
%%%%%%%%%%%%%%%%%%%%%%%%%%%%%%%%%%%%%%%%%%%%%%%%%%%%%%%%%%%%%%%%%%%%%%%%%%%%%%%%%%%%%%%%%%%%%%%%	
\title{The Casimir energy in Maxwell-Chern-Simons electrodynamics with various perfect electric or magnetic boundary condition combinations }
%%%%%%%%%%%%%%%%%%%%%%%%%%%%%%%%%%%%%%%%%%%%%%%%%%%%%%%%%%%%%%%%%%%%%%%%%%%%%%%%%%%%%%%%%%%%%%%%	

\author{Philipe De Fabritiis} 
\email{pdf321@cbpf.br} 
\affiliation{CBPF $-$ Centro Brasileiro de Pesquisas Físicas, Rua Dr. Xavier Sigaud 150, 22290-180, Rio de Janeiro, Brazil}
\affiliation{KU Leuven Campus Kortrijk-Kulak, Department of Physics,	Etienne Sabbelaan 53 bus 7657, 8500 Kortrijk, Belgium}

\author{Wellisson B. De Lima} 
\email{wellisson.b.lima@gmail.com} 
\affiliation{Rua Herodiano Barbosa 323, 37170-000, Boa Esperança, Brazil}

\author{David Dudal}
\email{david.dudal@kuleuven.be} 
\affiliation{KU Leuven Campus Kortrijk-Kulak, Department of Physics,	Etienne Sabbelaan 53 bus 7657, 8500 Kortrijk, Belgium}

%%%%%%%%%%%%%%%%%%%%%%%%%%%%%%%%%%%%%%%%%%%%%%%%%%%%%%%%%%%%%%%%%%%%%%%%%%%%%%%%%%%%%%%%%%%%%%%%	
\begin{abstract}
We investigate the Casimir effect between two parallel wires in $(2+1)D$ Maxwell-Chern-Simons (MCS) electrodynamics under different boundary conditions and separated by a distance $L$. Using a functional approach in which the boundary constraints are implemented by auxiliary fields localized on the wires, we integrate out the gauge field and derive an effective boundary theory whose functional determinant allows us to compute the Casimir energy. We consider perfect electric conductor (PEC), perfect magnetic conductor (PMC), and infinitely permeable (PMC2) boundary conditions. For two PEC wires, we recover the known MCS result and show that it coincides with the result obtained in the Abelian limit of the Curci-Ferrari (CF) model, implying that the Casimir effect in this case probes the photon mass $\mu$ but not its origin. The same energy is obtained for two PMC2 wires. In sharp contrast, the Casimir energy vanishes identically for the covariant PMC condition in MCS theory, both for PMC-PMC and PEC-PMC. Since this cancellation does not occur in the CF model, it could provide a signature of the topological origin of the photon mass in $(2+1)D$. Finally, the PEC-PMC2 case yields a non-vanishing Casimir force and reproduces the known result for a perfectly conducting wire facing an infinitely permeable one. Interestingly enough, we show that for PEC-PMC2, the Casimir force in the MCS theory is smaller than the one of the CF model for any $L,\mu >0$. Our results thus show that, depending on the boundary conditions, the Casimir effect could probe not only the photon mass but also reveal signatures of its origin.
\end{abstract}
	
\maketitle	
%%%%%%%%%%%%%%%%%%%%%%%%%%%%%%%%%%%%%%%%%%%%%%%%%%%%%%%%%%%%%%%%%%%%%%%%%%%%%%%%%%%%%%%%%%%%%%%%%%%%%%%%%%%%%%%%%%%%%%%%%%%%%%%%%%%%%%%%%%%%%%%%%%%%%%%%%%%%%%%%%%%%%%%%%%%%%%%%%%%%%%%%%%%%
\section{Introduction}

The Casimir effect~\cite{Casimir48} is one of the most striking manifestations of the quantum nature of the vacuum~\cite{Plunien86, Bordag01, Milton01, Milton04, Bordag09a}. Two neutral objects placed sufficiently close to each other can experience a measurable force, which can be understood as a consequence of the modification of the zero-point energy of quantum fields induced by the presence of nontrivial interfaces. One of the remarkable features of this phenomenon is the combination of conceptual depth and experimental accessibility. On the one hand, it is a genuine Quantum Field Theory (QFT) effect, rooted in the response of the vacuum to nontrivial boundary conditions and geometries. On the other hand, Casimir forces can be measured in low-energy laboratory experiments, without the need for particle accelerators~\cite{Lamoreaux97, Bimonte21}. This unusual combination has made the Casimir effect an exciting arena to investigate quantum vacuum effects.

A particularly interesting setting to explore the interplay between vacuum fluctuations and geometry is Maxwell-Chern-Simons (MCS) electrodynamics~\cite{Deser82a, Deser82b, Dunne99}, obtained by supplementing Maxwell theory in $(2+1)D$ with a Chern-Simons (CS) term. The latter can generate a gauge-invariant mass for the photon without introducing a Higgs field and without changing the number of physical propagating degrees of freedom. The CS term is topological in origin and odd under parity and time-reversal transformations.~\cite{Deser82a,Deser82b, Dunne99}. 

Beyond its field-theoretical interest, CS physics plays a central role in the effective description of condensed matter systems~\cite{Fradkin13, Witten16, Tong16, Witten26}.  In particular, CS theories arise naturally in the low-energy description of the fractional quantum Hall effect~\cite{Zhang89}, encoding flux attachment and the statistical transmutation underlying anyonic excitations~\cite{Wilczek82a, Wilczek82b}. They also appear in proposals for mechanisms of high-temperature superconductivity~\cite{Laughlin88}. This connection between topological field theory and experimentally relevant condensed matter phenomena makes MCS theory an attractive theoretical playground. 

The Casimir effect in MCS electrodynamics was first investigated by Milton and Ng~\cite{Milton90}, who considered two parallel perfectly conducting wires and found that the corresponding Casimir energy coincides with that of a massive scalar field~\cite{Ambjorn83}, with the topological photon mass playing the role of the scalar mass. The analysis was later extended to circular geometries~\cite{Milton92} and to other boundary conditions for parallel wires~\cite{Alves10}. The latter considered a perfectly conducting wire facing an infinitely permeable one and obtained a repulsive interaction, in analogy with the $(3+1)D$ result of Boyer~\cite{Boyer74}. They also showed that two infinitely permeable wires yield the same attractive interaction as two perfectly conducting wires~\cite{Alves10}. It is worth mentioning that there is a no-go theorem~\cite{Kenneth06} stating that to have a repulsive Casimir force, one needs to break the reflection symmetry of the system (see also~\cite{Bachas07, Asorey13}). This can be done using chiral materials~\cite{Jiang19}, topological insulators~\cite{Grushin11}, or Weyl semimetals~\cite{Wilson15}, for example. For the record, let us mention some other works on the Casimir effect for CS theories with boundaries~\cite{Bordag00, Bordag09b, Marachevsky23, Fialkovsky18, Vassilevich20}.

The Casimir effect is sensitive to the number of propagating degrees of freedom which are compatible with the boundary conditions and, in a massive theory, to the associated mass scale. But, could it also probe how that mass is generated? In $(2+1)D$, Maxwell theory contains one massless parity-preserving mode, whereas MCS electrodynamics contains one massive mode of definite helicity and explicitly violates parity. One may therefore ask whether the Casimir effect can give information about the topological and parity-odd origin of the photon mass, or whether it depends only on the existence of a mass gap. An useful comparison is provided by the Abelian limit of the Curci-Ferrari (CF) model~\cite{Dudal26a}, which also describes a massive photon, but without a CS term. Comparing the corresponding Casimir energies under different boundary conditions allows us to disentangle effects that depend only on the photon mass from those that are genuine fingerprints of its topological origin. Another interesting question is: does the parity-breaking provided by the CS term give a loophole for the no-go theorem~\cite{Kenneth06}? That is, considering a parity-odd electrodynamics would be enough to allow for a repulsive Casimir force even with a reflection-symmetric geometry?

The choice of boundary conditions is crucial to answer these questions. In this work we consider three gauge-invariant constraints representing distinct idealized interface responses. Perfect electric conductor (PEC) wires require the tangential electric field to vanish, as expected from a conductor in the limit of infinite conductivity. The perfect magnetic conductor (PMC) condition instead requires both the magnetic field and the perpendicular electric field to vanish. In $(3+1)D$ Maxwell electrodynamics, PMC arises naturally as the dual counterpart of PEC, reflecting the electromagnetic duality symmetry of the theory. Furthermore, it also has a phenomenological motivation coming from non-Abelian gauge theories through its connection with the dual superconductor picture of the Quantum Chromodynamics vacuum~\cite{Mandelstam76, Nambu74, tHooft81}. Finally, we consider the less restrictive condition in which only the magnetic field vanishes, conventionally associated with an idealized infinitely permeable wire. We denote this condition by PMC2 to distinguish it from the covariant PMC mentioned before. Although PMC and PMC2 differ only by the additional constraint on the normal electric field, they lead to strikingly different Casimir forces in MCS electrodynamics. Comparing them will be relevant to understand in which circumstances the Casimir force can be sensitive to the CS mass.

From a technical perspective, we employ the functional formulation of the Casimir effect developed in~\cite{Dudal20, Dudal22, Dudal23, Dudal24, Canfora25, Dudal26b, Dudal26a, Dudal26d}, whose original inspiration can be traced back to the boundary-QFT approach of Bordag, Robaschik, and Wieczorek~\cite{Bordag85}. In this framework, boundary conditions are imposed directly in the functional integral by introducing auxiliary fields living on the interfaces and acting as Lagrange multipliers. Integrating out the gauge field then leads to an effective action for the boundary fields, whose functional determinant encodes the Casimir energy. This framework provides a systematic, versatile, and efficient way of treating different types of boundary conditions. Moreover, this construction makes it particularly transparent which tensor structures of the bulk propagator survive in the effective boundary dynamics. This feature is especially relevant in MCS theory, where the propagator contains a parity-odd contribution associated with the CS term. Other complementary approaches to Casimir problems in gauge theories have also been developed, notably by Karabali and Nair~\cite{Karabali25, Karabali18, Karabali96}.

There are different ways of introducing boundaries in QFT. A more fundamental construction is provided by the Symanzik approach~\cite{Symanzik81}, in which admissible boundary conditions are derived from first principles, requiring properties such as locality, separability, and compatibility with the symmetries of the theory. This framework has been applied to several models closely related to the present one, including CS and MCS theories~\cite{Bertolini26, Blasi10, Maggiore18, Maggiore26}. Here we adopt instead a more phenomenological viewpoint, imposing boundary conditions to represent the (idealized) physical behavior of material interfaces. We are also not concerned with the compatibility of the interface boundary conditions with a variational principle, as we are still working on the usual $(2+1)D$ bulk space, for which all fields are assumed to drop off sufficiently fast at infinity. Our boundary conditions are thus not ``end of space'' conditions.

The main goal of this work is therefore to determine how the Casimir effect in a MCS electrodynamics depends on the boundary conditions and, in particular, whether it can distinguish a topological origin for the photon mass. We consider two parallel wires subject to PEC, PMC, and PMC2 boundary conditions, both in symmetric configurations (PEC-PEC, PMC-PMC, and PMC2-PMC2) and in the mixed PEC-PMC and PEC-PMC2 cases. Using a common functional framework, we recover previously known results for perfectly conducting and infinitely permeable wires~\cite{Milton90, Alves10} and extend the analysis to some cases which have not been studied before. The comparison with the Abelian limit of the CF model~\cite{Dudal26a} allows us to isolate the contribution associated with the CS term. As we shall show, some boundary conditions are sensitive only to the mass scale, whereas others are sensitive to its topological origin. 

This work is organized as follows. In Sec.~\ref{SecTheorSetup}, we introduce MCS electrodynamics and describe the geometry of two parallel wires under the PEC, PMC, and PMC2 boundary conditions used throughout the paper. We implement these conditions through auxiliary fields localized on the wires in Sec.~\ref{SecEffBoundAct}, deriving the corresponding effective boundary actions after integrating out the gauge field. The Casimir energy is computed in Sec.~\ref{SecCasimir} for a pair of wires satisfying the same boundary condition, namely PEC-PEC, PMC-PMC, and PMC2-PMC2. The results obtained until this point are discussed in Sec.~\ref{SecIntermezzo}, motivating the investigations that follow. In Sec.~\ref{SecMixedBC}, we extend the analysis to the mixed PEC-PMC and PEC-PMC2 cases. Finally, we summarize our main results in Sec.~\ref{SecConclusion}, gathering the final results for the Casimir energy under the different boundary conditions studied here in Table~\ref{TableSummary}. The detailed counting of classically propagating bulk degrees of freedom for each case is given in App.~\ref{AppCountingDOF}. An inequality relating the Casimir force in different limits for the PEC-PMC2 case is derived in App.~\ref{AppIneqMixedZeta}.

%%%%%%%%%%%%%%%%%%%%%%%%%%%%%%%%%%%%%%%%%%%%%%%%%%%%%%%%%%%%%%%%%%%%%%%%%%%%%%%%%%%%%%%%%%%%%%
\section{Theoretical setup}\label{SecTheorSetup}

\subsection{Maxwell-Chern-Simons theory}

Let us consider the Maxwell-Chern-Simons (MCS) action in a $(2+1)D$ Euclidean spacetime,
\begin{align}\label{eq:MCS}
    S_{\rm MCS} = \int \! d^3x \, \left[\frac{1}{4} F_{\mu \nu} F_{\mu\nu}-i~\frac{\mu}{2} ~ \epsilon_{\mu \nu \rho} A_\mu \partial_\nu A_\rho \right],
\end{align}
where $F_{\mu\nu}=\partial_\mu A_\nu-\partial_\nu A_\mu$ is the field strength, $\mu$ is the CS mass parameter, and $\epsilon_{\mu\nu\rho}$ is the $3D$ Levi-Civita symbol, with $\epsilon_{012}=+1$. The Euclidean indices are contracted with $\delta_{\mu \nu}$ and the factor of $i$ in the CS term comes from the Wick rotation. We take as mass dimension $\dim(A)=1/2$, so that indeed $\dim(\mu)=1$.

Before proceeding, let us briefly review the properties of MCS electrodynamics that will be important below (for an introductory review, see~\cite{Dunne99}). The addition of the CS term profoundly modifies the infrared content of Maxwell theory in three dimensions. Most importantly, it provides a mechanism for generating mass for the gauge field without introducing a Higgs field or explicitly breaking gauge invariance, since it is gauge invariant up to a surface term\footnote{In this work, we consider the wires as interfaces embedded in the unbounded Euclidean space $\mathbb{R}^3$. Assuming that the fields vanish in the asymptotic limit, we can safely neglect total derivatives.}. The resulting mass is usually referred to as a topological mass, reflecting the special topological character of the CS term, for example, by not contributing explicitly to the energy-momentum tensor. Here, the corresponding mass scale is set by $\mu$.

This can be seen directly from the classical equations of motion. Introducing the dual field strength as $\tilde{F}_\mu=\frac{i}{2}\epsilon_{\mu\nu\rho}F_{\nu\rho}$, one finds that the equations of motion $\partial_\nu F_{\nu\mu} + \mu \tilde{F}_\mu = 0$, together with the Bianchi identity, $\partial_\mu \tilde{F}_\mu = 0$, gives $\left(-\partial^2 + \mu^2\right) \tilde{F}_\mu = 0$, a massive wave equation for the gauge-invariant field $\tilde{F}_\mu$, with mass $\mu$. Despite being massive, the theory contains only one single physical propagating degree of freedom. This is a characteristic feature of topological mass generation: unlike the conventional Higgs mechanism, no additional degree of freedom is required to provide a longitudinal polarization for the photon. From the equations of motion, one can also see that a pure CS theory is non-propagating.

Another fundamental feature of MCS electrodynamics is its parity oddness. In $(2+1)D$, parity is defined as the reflection of one spatial coordinate. While the Maxwell term is invariant under this transformation, the CS term changes sign. The same term is also odd under time reversal. Consequently, a  MCS theory with a fixed non-zero value of $\mu$ explicitly breaks parity and time-reversal invariances. This is related to the fact that spin is a pseudo-scalar in $(2+1)D$, and a MCS theory describes one massive propagating degree of freedom with spin equal to $ {\rm sgn}(\mu)$. We remark that one can construct a parity-even MCS theory by appropriately adding another gauge field~\cite{Hagen92, Wilczek92, DelCima15, Well20, Well21, Well22, Well23}.

These properties are particularly relevant for the Casimir problem considered below. The mass $ \mu $ introduces a characteristic length scale $1/ \mu$ and the parity-odd term proportional to $\epsilon_{\mu\nu\rho}$ appearing in the MCS propagator carries information that is absent for a photon endowed with a non-topological mass. Depending on the boundary conditions, this information may either disappear from the effective boundary dynamics or leave a direct imprint on the Casimir interaction. This distinction will play a pivotal role in the results of this work.

Quantizing the theory through the functional integral formalism, one has to perform a gauge-fixing to find meaningful results. Using the Faddeev-Popov method, the action in linear covariant gauge can be written as
\begin{align} \label{ActionMCS}
    S_0 = S_{\rm MCS}  + \int \! d^3x \left(i h \, \partial_\mu A_\mu + \frac{\xi}{2} h^2\right),
\end{align}
where $\xi$ is a gauge parameter and $h$ is the Nakanishi-Lautrup field. The ghost sector decouples in this case, thus, we neglect it here\footnote{If the $z$-direction was compactified, the ghost would play a role, see~\cite{Dudal20}.}. Therefore, the functional integral governing the MCS theory in free space is
\begin{align}
    Z = \int \! \mathcal{D} A \mathcal{D} h \,\, e^{-S_0}.
\end{align}
Our goal is to investigate the vacuum energy of MCS electrodynamics in the presence of nontrivial interfaces. We now introduce the geometry and the boundary conditions considered throughout this work.

\subsection{Geometry and boundary conditions}

Let us consider a pair of parallel wires, infinitely thin and large, separated by a distance $L$ from each other, located at $y=y_\pm \equiv \pm L/2$. In the following, $\Sigma^\pm$ are the $2D$ subspaces defined by the wires at $y = y_\pm$ and $n_\mu = \left(0,0,1\right)$ is the unit vector normal to the wires\footnote{To be precise, the outward normal should be $\pm \delta_{\mu y}$ for the wires $\Sigma^\pm$, but since this sign will not play any role here, we neglect it.}. Greek indices such as $\mu, \nu$ will run over $\{t,x,y\}$; Latin indices such as $i, j$ over $\{t,x\}$. $\bold{k}$ denotes the vector $\left(k_t, k_x\right)$.

We are interested in perfect electric conductor (PEC), perfect magnetic conductor (PMC), and infinitely permeable (PMC2) boundary conditions,  defined by
\begin{align} \label{BoundaryConditions}
     \tilde{F}_\mu n_\mu \big\vert_{\Sigma_\pm} &= 0 \quad {(\rm PEC)}, \nonumber \\
     F_{\mu \nu} n_\nu \big\vert_{\Sigma_\pm} &= 0 \quad  {(\rm PMC)}, \nonumber \\
     F_{x \nu} n_\nu \big\vert_{\Sigma_\pm} &= 0 \quad  {(\rm PMC2)},
\end{align}
where we defined the dual as $\tilde{F}_\mu \equiv \frac{i}{2}\epsilon_{\mu \nu \rho} F_{\nu \rho}$. Considering $n_\mu = \left(0,0,1\right)$, these boundary conditions mean that in the wires we have: $E_x = 0$ (PEC); $E_y = B = 0$ (PMC); $B = 0$ (PMC2), where we defined the electric and magnetic fields as $E_i \equiv F_{0i}$ and $B \equiv F_{xy}$. Since these conditions are written in terms of the field strength, they are gauge invariant. 

For notational convenience, we define the following objects: $G \equiv \tilde{F}_\mu n_\mu$ and $G_\mu \equiv F_{\mu \nu} n_\nu$ (note that $G_y \equiv 0$). Then, we can write the boundary conditions as $G \vert_{\Sigma_\pm} = 0$ for PEC, $G_\mu \vert_{\Sigma_\pm} = 0$ for PMC, and $G_x \vert_{\Sigma_\pm} = 0$ for PMC2. 

One can impose the boundary conditions described above directly in the functional integral by introducing suitable auxiliary fields $b^\pm$, $b_i^\pm$, and $b_x^\pm$, which can be included in the action through the addition of the terms
\begin{align}
S_{\rm BC} \!=\! 
    \begin{cases}
    \! \int_{\Sigma^-} \! d^2 \bold{x} \, b^{-} (\bold{x}) G+ \int_{\Sigma^+} \! d^2 \bold{x} \, b^{+} (\bold{x}) G, \!\!\! &\text{(PEC)},\\
    \! \int_{\Sigma^-} \! d^2 \bold{x} \, b_i^{-} (\bold{x}) G_i+ \int_{\Sigma^+} \! d^2 \bold{x} \, b_i^{+} (\bold{x}) G_i, \!\!\! &\text{(PMC)},\\
    \! \int_{\Sigma^-} \! d^2 \bold{x} \, b_x^{-} (\bold{x}) G_x \!+\! \int_{\Sigma^+} \! d^2 \bold{x} \, b_x^{+} (\bold{x}) G_x, \!\!\! &\text{(PMC2)}.
    \end{cases}
\end{align}
The auxiliary fields $b^\pm, b_i^\pm, b_x^\pm$ only depend on $\bold{x} = (t,x)$ and act as Lagrange multipliers. The boundary conditions can be enforced performing a variation on them.

At this point one can already see that the PMC2 condition can be obtained from the PMC one by simply imposing $b_t^\pm = 0$ to throw away the constraint $G_t\vert_{\Sigma^\pm} = 0$. Thus, in what follows, we will stop writing explicit expressions for the PMC2 case to avoid cluttering the text, simply extracting them from the PMC case in the end. 

We can rewrite the boundary term in a more convenient way. Indeed, one can rewrite the action imposing the PEC and PMC boundary conditions respectively as
\begin{align}
    b^\gamma G = b^\gamma \tilde{F}_\rho n_\rho = \frac{1}{2} \left(i\epsilon_{\rho\mu\nu} n_\rho b^\gamma\right) F_{\mu \nu},
\end{align}
and
\begin{align}
    b_i^\gamma G_i = b_\mu^\gamma F_{\mu \nu} n_\nu = \frac{1}{2} \left(b_\mu^\gamma n_\nu - b_\nu^\gamma n_\mu\right) F_{\mu \nu},
\end{align}
in which we have used $b_y^\gamma \equiv 0$ and we have $\gamma \in \{+,-\}$. Therefore, the action  $S_{\rm BC}$ can be simply rewritten as
\begin{align}\label{Action_BC}
S_{\rm BC} = \int \! d^3x \, \frac{1}{2} H_{\mu \nu} F_{\mu \nu},
\end{align}
with the antisymmetric tensor $H_{\mu \nu}$ defined by 
\begin{align} \label{HtensorConfig}
H_{\mu\nu}(x) = \delta(y - y^\gamma) \, h_{\mu \nu}^\gamma(\bold{x}),
\end{align}
where the plate indices $\gamma \in \{+,-\}$ are summed and the definition of $h_{\mu \nu}^\gamma$ depends on the boundary conditions,
\begin{align} \label{htensor}
h_{\mu\nu}^\gamma = 
    \begin{cases}
    i \epsilon_{\mu\nu\rho} n_\rho b^\gamma, \quad &\text{(PEC)},\\
    b_\mu^\gamma n_\nu - b_\nu^\gamma n_\mu, \quad &\text{(PMC)}.
    \end{cases}
\end{align}
Here we are assuming that $b_y^\pm \equiv 0$. As already emphasized, the PMC2 case is not explicitly written because it can be suitably recovered from the PMC result after.

Summing the MCS action with the $S_{\rm BC}$ written above and integrating out the Nakanishi-Lautrup field $h$, we will find the full action for a MCS theory in the presence of parallel wires under the boundary conditions~\eqref{BoundaryConditions}:
\begin{align}\label{Action_MCSplusBC}
    S = \! \int \!\! d^3x &\left( \frac{1}{4} F_{\mu \nu} F_{\mu\nu} -i~\frac{\mu}{2} ~ \epsilon_{\mu \nu \rho} A_\mu \partial_\nu A_\rho \right. \nonumber \\&\left.+ \frac{1}{2 \xi} \left(\partial_\mu A_\mu\right)^2  + \frac{1}{2}  H_{\mu \nu} F_{\mu \nu}\right).
\end{align}
The functional integral associated with this action is
\begin{align}
   Z =  \int \! \mathcal{D}A \, \mathcal{D} b^+ \mathcal{D} b^- \,\, e^{-S}.
\end{align}

Having the partition function for this system, we could integrate out the boundary fields to obtain an effective action for the photon~\cite{Bordag85}. However, we choose to follow another path. Indeed, following the procedure adopted in~\cite{Dudal20, Dudal22, Dudal23, Dudal24, Canfora25, Dudal26a, Dudal26b, Dudal26d}, we will integrate out the photon to obtain an effective boundary action describing the dynamics of the auxiliary fields $b^\pm$, $b_i^\pm$. This 2D effective boundary theory will contain the relevant information about our system and allow us to compute the Casimir energy. 

%%%%%%%%%%%%%%%%%%%%%%%%%%%%%%%%%%%%%%%%%%%%%%%%%%%%%%%%%%%%%%%%%%%%%%%%%%%%%%%%%%%%%%%%%%%%%%
\section{The effective boundary action}\label{SecEffBoundAct}

The full action~\eqref{Action_MCSplusBC} can be written in Fourier space as
\begin{align} \label{Action_MCS_Fourier}
    S \!= \frac{1}{2} \!\int \!\!\! \frac{d^3k}{(2 \pi)^3} \bigg\lbrace  A_\mu(k) K_{\mu \nu} A_\nu(-k) + v_\nu(k) A_\nu(-k) \bigg\rbrace,
\end{align}
where the photon quadratic operator is given by
\begin{align}
    K_{\mu \nu} = k^2 \left[\delta_{\mu \nu} - \left(1 - \frac{1}{\xi}\right) \frac{k_\mu k_\nu}{k^2} +\frac{\mu}{k^2}\epsilon_{\mu\rho\nu}k_\rho\right],
\end{align}
and we have also defined
\begin{align}
    v_\nu(k) \equiv 2 i k_\mu H_{\mu \nu}(k).
\end{align}
The operator $K_{\mu \nu}$ has a well-defined inverse, given by 
\begin{align}
    \left(K^{-1}\right)_{\mu \nu} = \frac{1}{k^2+\mu^2} &\left[\delta_{\mu \nu} + \left(\xi - 1 + \xi\frac{\mu^2}{k^2}\right) \frac{k_\mu k_\nu}{k^2}\right. \nonumber\\ &\left.- \zeta\frac{\mu}{k^2}\epsilon_{\mu\rho\nu}k_\rho\right].
\end{align}
Notice that we included a label $\zeta$ in the above expression. In this way, we can keep track of the CS contribution, being able to immediately find the result for the Abelian CF model by taking the limit $\zeta \to 0$ in the end. Taking $\mu \to 0$ we obtain Maxwell\footnote{There is a subtlety here. In some cases, one can obtain different results depending on when the limit $\mu \to 0$ is taken. This can be related with the emergence of a boundary gauge symmetry in such cases, as we will see later (see~\cite{Dudal26a} for more details).}. Thus, our approach will allow us to scrutinize these three scenarios with a single effort. However, it must be emphasized that this is just a practical way to compare the results, rather than a physically meaningful limit.

From the structure of Eq.~\eqref{Action_MCS_Fourier} one can readily see that we can decouple the photon from the auxiliary fields by performing a suitable shift in $A_\mu$. Indeed, performing 
\begin{align}
A_\mu(k) \rightarrow A_\mu(k) - \frac{1}{2} v_\rho(k) \left(K^{-1}\right)_{ \rho\mu}(k),    
\end{align}
we will find for the action 
\begin{align}
    S = \frac{1}{2} \int \! \frac{d^3k}{(2\pi)^3} &\bigg\lbrace A_\mu(k) K_{\mu \nu} A_\nu(-k) \nonumber \\
    &- \frac{1}{4} v_\mu(k) \left(K^{-1}\right)_{\mu \nu} v_\nu(-k) \bigg\rbrace.
\end{align}
Being a mere shift in the field $A_\mu$ by a quantity that does not depend on the fields itself, we can readily conclude that the functional integral measure will be left invariant.

The above expression clearly shows that the photon decouples from the auxiliary fields in the functional integral. That is, we can write $Z = Z_A Z_b$. Since the photon part, after the field redefinition, does not have any information about the wires separation, it does not know about the geometry, and thus will not contribute to the Casimir effect. Thus, we can focus on the boundary field functional integral, which will contain the relevant information about the gauge sector. Therefore, we consider
\begin{align}
    Z_b = \int \mathcal{D}b^- \mathcal{D}b^+ e^{-S_b},
\end{align}
with the effective boundary action defined as 
\begin{align} \label{BoundEffAct1}
    S_b = -\frac{1}{2} \! \int \!\!\frac{d^3k}{(2\pi)^3} \left[ k_\mu H_{\mu \rho}(k) \left(K^{-1}\right)_{\rho \sigma} k_\nu H_{\nu \sigma}(-k) \right].
\end{align}
According to Eq.~\eqref{HtensorConfig} we have $H_{\mu\nu}(x) = \delta(y - y^\gamma) h_{\mu \nu}^\gamma(\bold{x})$. Thus, in Fourier space, we find 
\begin{align}
    H_{\mu \nu}(k) = e^{i k_y y^\gamma} h_{\mu \nu}^{\gamma}(\bold{k}).
\end{align}
Plugging this expression into the effective boundary action, we find
\begin{align}
    S_b = -&\frac12 \int \!\!\frac{d^2 \bold{k}}{(2\pi)^2}  h_{\mu \rho}^{\gamma}(\bold{k}) h_{\nu \sigma}^{\lambda}(-\bold{k}) \nonumber \\
    &\times \int \! \frac{dk}{2\pi} \, e^{i k_y \left(y_\gamma - y_\lambda\right)} k_\mu k_\nu \left(K^{-1}\right)_{\rho \sigma}.
\end{align}
This expression is quadratic in the boundary fields, but it is written in terms of the $h_{\mu \nu}^\gamma$ tensor, which is different for the PEC and PMC boundary conditions~\eqref{htensor}. The next step is to obtain the effective action $S_b$ written as
\begin{align}
    S_b = \frac{1}{2} \int \! \frac{d^2 \bold{k}}{(2\pi)^2} \, b(\bold{k}) \,\mathbb{K}(\bold{k})  \, b(-\bold{k}). 
\end{align}
Naturally, the boundary dynamical operator $\mathbb{K}$ will depend on the condition adopted. Therefore, we will now consider the different boundary conditions separately. 

%%%%%%%%%%%%%%%%%%%%%%%%%%%%%%%%%%%%%%%%%%%%%%%
\subsection{PEC wires}

For PEC boundary conditions, $h_{\mu\nu}^\gamma = i \epsilon_{\mu\nu\rho} n_\rho b^\gamma$. Thus, the effective boundary action reduces to
\begin{align}
    S_b = \frac{1}{2} \int &\frac{d^3k}{(2\pi)^3} \left[ e^{i k_y (y^\gamma - y^\lambda)}  b^{\gamma}(\bold{k}) b^{\lambda}(\bold{-k}) \right. \nonumber \\ 
    &\left. \times k_\mu k_\nu \epsilon_{\mu\rho\alpha}n_\alpha\epsilon_{\nu\sigma\beta}n_\beta\left(K^{-1}\right)_{\rho \sigma}  \right].
\end{align}
Now we discuss the contraction present in the second line. The contribution proportional to $k_\rho k_\sigma$ in $K^{-1}$ will vanish due to the antisymmetry of the $\epsilon$ symbol, thus the gauge parameter $\xi$ disappears automatically. The contribution proportional to $\epsilon_{\rho\tau\sigma}k_\tau$ also vanishes, after one makes use of the identity $\epsilon_{\rho\tau\sigma}\epsilon_{\rho\mu\alpha} = \delta_{\tau\mu}\delta_{\sigma\alpha} - \delta_{\tau\alpha}\delta_{\sigma\mu}$ Thus,
\begin{align}
    X &\equiv k_\mu k_\nu \epsilon_{\mu\rho\alpha}n_\alpha\epsilon_{\nu\sigma\beta}n_\beta\left(K^{-1}\right)_{\rho \sigma} \nonumber \\
    &= \frac{1}{k^2+\mu^2}k_\mu k_\nu n_\alpha n_\beta \epsilon_{\mu\rho\alpha} \epsilon_{\nu\rho\beta}\nonumber \\ 
    &=\frac{1}{k^2+\mu^2} n_\alpha n_\beta (k^2  \delta_{\alpha\beta} - k_\alpha k_\beta)\nonumber \\
    &= \frac{\vert\bold{k}\vert^2}{k^2+\mu^2},
\end{align}
where $\vert\bold{k}\vert^2 = k_t^2 + k_x^2$. Inserting the above contraction in the action $S_b$, we get
\begin{align}
    S_b = \frac{1}{2} &\int \!  \frac{d^2\bold{k}}{(2\pi)^2} \bigg[ b^{\gamma}(\bold{k})\, \vert\bold{k}\vert^2 \, b^{\lambda}(-\bold{k}) \nonumber \\
    &\times\int \! \frac{dk_y}{2\pi}     \frac{e^{i k_y(y^\gamma- y^\lambda)}}{k^2+\mu^2} 
 \bigg]
\end{align}
The next step is to perform the $dk_y$ integral. We need to use the following expression:
\begin{align}
    \int \! \frac{dk_y}{2\pi} \frac{e^{i k_y (y^\gamma - y^\lambda)}}{k^2+ \mu^2}  = \frac{1}{2 \sqrt{ \vert\bold{k}\vert^2 +\mu^2}} e^{- \sqrt{ \vert\bold{k}\vert^2 +\mu^2} \, \vert y^\gamma - y^\lambda \vert}.
\end{align}
Therefore, we find for the effective boundary action:
\begin{align}
    S_b = \frac{1}{2} \int \! \frac{d^2 \bold{k}}{(2\pi)^3} \, b^{\gamma}(\bold{k}) \,\mathbb{K}^{\gamma \lambda}(\bold{k})  \, b^{\lambda}(-\bold{k}),
\end{align}
where the boundary dynamical operator $\mathbb{K}$ is given by
\begin{align} \label{MCSKopPEC}
    \mathbb{K}^{\gamma \lambda}(\bold{k}) =  \frac12 \frac{\vert \bold{k} \vert^2}{ \sqrt{\vert \bold{k}\vert^2 + \mu^2}}   e^{- \sqrt{\vert \bold{k}\vert^2 + \mu^2} \, \vert y^\gamma - y^\lambda \vert}.
\end{align}
This agrees with the boundary dynamical  operator obtained for parallel wires in the Abelian limit of the CF model under PEC conditions~\cite{Dudal26a}\footnote{Although the expressions reported in~\cite{Dudal26a} for $\mathbb{K}^{\rm PEC}$ and its determinant have typos, the Casimir energy obtained is correct.}. Therefore, for PEC wires, we can already conclude that the relevant feature is that the photon has a mass, independently of its origin.

%%%%%%%%%%%%%%%%%%%%%%%%%%%%%%%%%%%%%%%%%%%%%%%%%%%%%%%%%%%%%%%%%%%%%%%%%%%%%%%%%%%%%%%%%%%%%%
\subsection{PMC wires}

Now, we will perform the analogous computation done above, but for PMC boundary conditions. In this case, we have $h_{\mu \nu}^\gamma = b_\mu^\gamma n_\nu - b_\nu^\gamma n_\mu$. Here, to factorize the b-fields, it will be convenient to introduce the tensor 
    \begin{align}
    h_{i \mu \nu} \equiv \delta_{i \mu} n_\nu - \delta_{i \nu} n_\mu,
\end{align}
such that $h_{\mu \nu}^\gamma = h_{i \mu \nu} b_i^\gamma$. The effective boundary action is
\begin{align}
    S_b = -\frac{1}{2} \int &\frac{d^3k}{(2\pi)^3} \left[ e^{i k_y (y^\gamma - y^\lambda)}  b_i^{\gamma}(\bold{k}) b_j^{\lambda}(\bold{-k}) \right. \nonumber \\ 
    &\left. \times k_\mu k_\nu h_{i \mu \rho} h_{j \nu \sigma}\left(K^{-1}\right)_{\rho \sigma}  \right].
\end{align}
As before, we discuss the contraction in the second line. The term with $k_\rho k_\sigma$ in $K^{-1}$ vanishes due to the antisymmetry of the boundary tensors $h_{i\mu \rho}$ and $h_{j\nu \sigma}$, considering the factor $k_\mu k_\nu$. Thus, the term with the gauge parameter $\xi$ will vanish,  as happened in the PEC case. There are two contributions coming from this contraction: one with $\delta_{\rho \sigma}$, that is already present in the Maxwell case; another one with $\epsilon_{\rho \tau \sigma}$, proportional to $\mu$. We have,
\begin{align}
    &Y_{ij} \equiv k_\mu k_\nu h_{i \mu \rho} h_{j \nu \sigma} \left(K^{-1}\right)_{\rho \sigma}   \nonumber \\
    &=\frac{k_\mu k_\nu}{k^2 + \mu^2} \left[\delta_{i \mu} \delta_{j \nu} - \delta_{j \nu} n_\mu n_i - \delta_{i \mu} n_\nu n_j + \delta_{ij} n_\mu n_\nu\right]  \nonumber \\
    &-\! \zeta\frac{\mu}{k^2}\frac{k_\mu k_\nu k_\tau}{k^2 + \mu^2} \epsilon_{\rho \tau \sigma} \! \left(\delta_{i \mu} n_\rho - \delta_{i \rho} n_\mu\right) \! \left(\delta_{j \nu} n_\sigma - \delta_{j \sigma} n_\nu\right).
\end{align}
After a tedious but straightforward computation, we find
\begin{align}
     Y_{ij} &= \frac{1}{k^2 + \mu^2} \left(k_i k_j + \delta_{ij} k_y^2\right) \nonumber \\
    &- \zeta\frac{\mu}{k^2}\frac{k_y}{k^2 + \mu^2}  \left(- k_\ell k_i \epsilon_{y \ell j} - k_\ell k_j \epsilon_{i \ell y} + k_y^2 \epsilon_{i y j}  \right).
\end{align}
In the above expression, the first line is coming from the $\delta_{\rho \sigma}$ contribution, being exactly what we would obtain in the Abelian limit of the CF model (for Maxwell, we would need only to change the pre-factor to $1/k^2$). The second line is the footprint of the CS term, which will bring the novel features of this model. Using symmetry arguments and defining $\epsilon_{ij} \equiv\epsilon_{yij}$, one can show that
\begin{align}
    - k_\ell k_i \epsilon_{y \ell j} - k_\ell k_j \epsilon_{i \ell y} = - \vert\bold{k}\vert^2 \epsilon_{ij}.
\end{align}
This allows us to rewrite the second line in a simpler form. Therefore, for the contraction $Y_{ij}$, we can write
\begin{align}
    Y_{ij} &= \frac{1}{k^2 + \mu^2} \left[k^2 \delta_{ij} - \bold k^2 \Theta_{ij}(\bold k) + \zeta \mu k_y \epsilon_{ij}\right],
\end{align}
where $\Theta_{ij} \equiv \delta_{ij} - \Omega_{ij}$ and $\Omega_{ij} \equiv k_i k_j / \vert\bold{k}\vert^2$.
Inserting this information into $S_b$, we find 
\begin{align}
    &S_b = -\frac{1}{2} \int \!  \frac{d^2\bold{k}}{(2\pi)^2} \bigg\lbrace b_i^{\gamma}(\bold{k})  \, b_j^{\lambda}(-\bold{k}) \nonumber \\
    &\int \!\! \frac{dk_y}{2\pi} \! \left[\! \frac{e^{i k_y(y^\gamma- y^\lambda)}}{k^2+\mu^2} \! \left( k^2 \delta_{ij} \!- \vert\bold k\vert^2 \Theta_{ij}(\bold k) \!+ \zeta \mu k_y \epsilon_{ij}\right) \! \right] \!
 \bigg\rbrace. 
\end{align}
To perform the integrals in $dk_y$, we use the results:
\begin{align}
    \int \! \frac{dk_y}{2 \pi} \frac{e^{i k_y (y^\gamma- y^\lambda)}}{k^2 + \mu^2} &= \frac{e^{-\omega \vert y^\gamma- y^\lambda \vert}}{2 \omega}, \nonumber \\
    \int \!\! \frac{dk_y}{2 \pi} \frac{k_y \, e^{i k_y (y^\gamma- y^\lambda)}}{k^2 + \mu^2} &= \frac{i}{2} {\rm sgn}(y^\gamma- y^\lambda) e^{-\omega \vert y^\gamma- y^\lambda \vert}, \nonumber \\
     \int \! \frac{dk_y}{2 \pi} \frac{k_y^2 e^{i k_y (y^\gamma- y^\lambda)}}{k^2 + \mu^2} &= \delta\left(y^\gamma - y^\lambda\right) -\frac{\omega}{2} e^{-\omega \vert y^\gamma- y^\lambda \vert}, 
\end{align}
where $\omega \equiv \sqrt{\vert\bold{k}\vert^2 + \mu^2}$ and ${\rm sgn}(x) \equiv x / \vert x \vert$ is the sign function. We discard the $\delta(y^\gamma - y^\lambda)$ appearing in the last integral above since it gives us either $0$ or $\delta(0)$, which can be safely neglected using dimensional regularization~\cite{Collins84} (see also the corresponding discussion in~\cite{Dudal26c}). Thus, the effective boundary action for PMC plates reads
\begin{align} \label{SbPMC}
    S_b = \frac{1}{2} \int \! \frac{d^2 \bold{k}}{(2\pi)^2} \, b_i^{\gamma} \,\mathbb{K}_{ij}^{\gamma \lambda}(\bold{k})  \, b_j^{\lambda},
\end{align}
where the boundary dynamical operator $\mathbb{K}$ is given by
\begin{align} \label{MCSKopPMC}
&\mathbb{K}_{ij}^{\gamma \lambda}(\bold{k}) = \frac{\sqrt{\vert\bold{k}\vert^2 + \mu^2}}{2}\left[ \Theta_{ij} + \left(\frac{\mu^2}{\vert\bold{k}\vert^2 + \mu^2}\right) \Omega_{ij} \right. \nonumber \\
&\left. - i \zeta \left(\frac{\mu}{\sqrt{\vert\bold{k}\vert^2 + \mu^2}}\right) \epsilon_{ij} {\rm sgn}(y^\gamma- y^\lambda) \!\right]\! e^{-\sqrt{\vert\bold{k}\vert^2 + \mu^2} \vert y^\gamma- y^\lambda \vert},
\end{align}
Discarding the CS term contribution taking $\zeta = 0$, the expression perfectly matches the one obtained in~\cite{Dudal26a}, for PMC wires in the Abelian limit of the CF model. Thus, we have already a hint that for PMC wires, the CS origin of the photon mass can leave its fingerprints and give a result different from the Abelian CF model one.

%%%%%%%%%%%%%%%%%%%%%%%%%%%%%%%%%%%%%%%%%%%%%%%%%%%%
\subsection{PMC2 wires}

The PMC2 case can be extracted directly from the PMC result by just imposing $b_t^\gamma = 0$, as already highlighted. That is, if one starts with the PMC2 boundary condition and repeat all the steps, one would find a similar result for the effective boundary action~\eqref{SbPMC}, but with only the $b_x$ auxiliary field, and a dynamical operator $\mathbb{K}_{xx}^{\gamma \lambda}$ that can be obtained directly from Eq.~\eqref{MCSKopPMC} by putting $i=j=x$. Doing so, one obtains the following result:
\begin{align}  \label{MCSKopPMC2}
     \mathbb{K}_{xx}^{\gamma \lambda}(\bold{k}) = \frac{1}{2}\frac{k_t^2 + \mu^2}{\sqrt{\bold{\vert k\vert}^2 + \mu^2}} e^{-\sqrt{\bold {\vert k\vert}^2 + \mu^2}\vert y^\gamma - y^\lambda\vert}.
\end{align}
Notice that the $\zeta$-dependence disappeared. This means that the results coming from MCS will be the same as the ones coming from the Abelian limit of the CF model.

%%%%%%%%%%%%%%%%%%%%%%%%%%%%%%%%%%%%%%%%%%%%%%%%%%%%%%%%%%%%%%%%%%%%%%%%%%%%%%%%%%%%%%%%
\section{Casimir energy} \label{SecCasimir}

The vacuum energy can be computed in the functional integral approach through the following expression\footnote{For a pedagogical derivation of the method employed to compute the
Casimir energy from the partition function, see~\cite{Dudal26b}.}:
\begin{align} \label{ZexpVE}
     Z = e^{- V \mathcal{E}},
 \end{align}
where $\mathcal{E}$ is the vacuum energy density and $V$ is the volume of spacetime. The existence of a non-trivial interface such as parallel wires modify the vacuum energy, which becomes dependent on the wires separation. The Casimir energy can be defined as the difference between the vacuum energy of the system in the presence of wires and in their absence. In practice, the latter can be obtained from the former by taking the limit $L \to \infty$. Therefore, when computing the Casimir energy, only terms depending on $L$ will contribute, since all the others will cancel when we subtract the vacuum energy of the free space.

The functional integral of our system has factorized into two independent contributions: a purely photonic $Z_A$ and a boundary one $Z_b$. Since the photonic part does not have any $L$-dependence after the shift, it does not contribute to the Casimir energy. Therefore, we can safely restrict our attention to $Z_b$. Now, since $S_b$ is quadratic in the boundary fields, one can compute $Z_b$ by performing a Gaussian functional integral, that is,
\begin{align}
    Z_b = \int \! \mathcal{D}b^- \mathcal{D}b^+ e^{-S_b} = C \left( {\rm Det}  \mathbb{K} \right)^{-1/2},
\end{align}
where $C$ is a constant infinite factor and $\mathbb{K}$ is the boundary dynamical operator, which is given in Eqs.~\eqref{MCSKopPEC},\eqref{MCSKopPMC}, and~\eqref{MCSKopPMC2} depending on the boundary condition. The determinant above must be understood in the functional sense ({\it i.e.}, both as a matrix determinant on the discrete indices and the integral in the continuous index $\bold{k}$). 

Using the identity $\log {\rm Det} A= {\rm Tr} \log A$ with respect to the continuous index and the expression relating the partition function to the vacuum energy~\eqref{ZexpVE}, we can find
\begin{align} \label{VacuumEnergy}
    \mathcal{E} = \frac{1}{2} \int \frac{d^2 \bold{k}}{(2\pi)^2} \log \left(\det \mathbb{K}(\bold{k})\right),
\end{align}
where $\det \mathbb{K}(\bold{k})$ refers only to the matrix determinant, which we need to compute now. Since the boundary dynamical operator is different for each case, we will do this computation separately in the following. 

%%%%%%%%%%%%%%%%%%%%%%%%%%%%%%%%%%%%%%%%%%%%%%%
\subsection{PEC wires}

For PEC boundary conditions, as one can see in Eq.~\eqref{MCSKopPEC}, the boundary dynamical operator $\mathbb{K}^{\gamma \lambda}$ is a $2 \times 2$ matrix in the discrete indices, which can be written as
\begin{align}
\mathbb{K}^{\rm PEC} = 
\begin{pmatrix}
    \mathbb{K}^{++} & \mathbb{K}^{+-} \\
    \mathbb{K}^{-+} & \mathbb{K}^{--} 
\end{pmatrix}.
\end{align}
It is immediate to compute its determinant. We get
\begin{align}
    \det \mathbb{K}(\bold{k}) = \frac{1}{4} \frac{\vert \bold{k} \vert^4}{\vert \bold{k} \vert^2 + \mu^2} \left(1 - e^{-2 L \sqrt{\vert \bold{k}\vert^2 +\mu^2}} \right).
\end{align}
We can then insert this expression into the vacuum energy formula~\eqref{VacuumEnergy}. However, as already argued, only the $L$-dependent factors are relevant for the Casimir energy. Thus, discarding $L$-independent factors, we get
\begin{align}
    \mathcal{E}_{\rm PEC} &= \frac{1}{2} \int \! \frac{d^2 \bold{k}}{(2\pi)^2} \log\left[1 - e^{-2 L \sqrt{\vert \bold{k}\vert^2 +\mu^2}}  \right] \nonumber \\
   % &=  \frac{1}{4\pi} \int_0^{\infty} \! dk \, k  \log\left[1 - e^{-2 L \sqrt{k^2 +\mu^2}}  \right] \nonumber \\
    &=  \frac{1}{4\pi}\! \!\int_{\mu}^{\infty} \! d\omega \, \omega  \log\left[1 - e^{-2 L \omega}  \right].
\end{align}
Computing this integral, we can find the final result:
\begin{align}
       \mathcal{E}_{\rm PEC} &= -\frac{1}{16 \pi L^2} \left[2 L \mu \, {\rm Li}_2\left(e^{-2L \mu}\right) + {\rm Li}_3\left(e^{-2L \mu}\right)\right],
\end{align}
where ${\rm Li}_n(z) = \sum_{k=1}^{\infty} \frac{z^k}{k^n}$ is the polylogarithm function. This is exactly the result obtained in~\cite{Dudal26a}, as expected due to the boundary dynamical operator form. 

Furthermore, our result agrees with the one obtained in~\cite{Milton90}, where the authors computed the Casimir force of two perfectly conducting wires in a MCS electrodynamics through the energy-momentum tensor using Green's functions techniques. Indeed, they obtained:
\begin{align}
    f_{\rm PEC} = - \frac{1}{16 \pi L^3} \int_{2 L \mu}^{\infty} \! dy \, \frac{y^2}{e^{y} - 1}.
\end{align}
This integral can be computed analytically using {\it Mathematica} in terms of the polylogarithm functions, giving
\begin{align}
    f_{\rm PEC} = \frac{1}{8 \pi L^3} &\left[2 L^2\mu^2 \log\left(1 - e^{-2 L \mu}\right) \right.\nonumber \\
    &\left. -2 L \mu \, {\rm Li}_2\left(e^{-2 L \mu}\right) - {\rm Li}_3\left(e^{-2 L \mu}\right)\right].
\end{align}
This perfectly agrees with our result by taking $f = -\frac{\partial \mathcal{E}}{\partial L}$. 

Now, taking the Maxwell limit (that is, $\mu \to 0$), we find
\begin{align}
    \lim_{\mu \to 0} \mathcal{E}_{\rm PEC} = -\frac{\zeta(3)}{16 \pi L^2}, 
\end{align}
which is the known result for the Casimir energy of two perfectly conducting parallel wires in $3D$, see e.g.~\cite{Wolfram83}.

Thus, we obtained a Casimir energy for PEC wires which agrees with the known result obtained through other methods~\cite{Milton90, Wolfram83}. Furthermore, this result matches with the Abelian limit of the CF model~\cite{Dudal26a}. That is, the Casimir energy for PEC wires can ``feel'' the mass of the photon, but it cannot distinguish its origin as arising from a CS term, or not. The story for the PMC case is strikingly different, as we discuss below.

%%%%%%%%%%%%%%%%%%%%%%%%%%%%%%%%%%%%%%%%%%%%%%%
\subsection{PMC wires}

For PMC boundary conditions, considering  Eq.~\eqref{MCSKopPMC}, the boundary dynamical operator $\mathbb{K}^{\gamma \lambda}_{ij}$ is a $4 \times 4$ matrix in the discrete indices, composed by four $2 \times 2$ blocks:
\begin{align}
\mathbb{K}_{\rm PMC} = 
\begin{pmatrix}
    \mathbb{K}_{ij}^{++} & \mathbb{K}_{ij}^{+-} \\
    \mathbb{K}_{ij}^{-+} & \mathbb{K}_{ij}^{--} 
\end{pmatrix}.
\end{align}
The determinant of a block matrix can be computed as 
\begin{align}\label{detblock}
    \det 
    \begin{pmatrix}
    A & B \\
    C & D 
\end{pmatrix} = \det \left[A - B D^{-1} C\right] \, \det [D],
\end{align}
under the assumption that $D^{-1}$ exists, naturally. Writing the blocks explicitly for the reader's convenience, we have
\begin{align}
    A &\equiv \mathbb{K}_{ij}^{++} = \frac{\omega}{2} \left[\Theta_{ij} + \left(\frac{\mu^2}{\omega^2}\right) \Omega_{ij} \right], \nonumber \\
    B &\equiv \mathbb{K}_{ij}^{+-} = \frac{\omega}{2} \left[\Theta_{ij} + \left(\frac{\mu^2}{\omega^2}\right) \Omega_{ij} -  i \zeta \left(\frac{\mu}{\omega}\right) \epsilon_{ij} \right] e^{-\omega L}, \nonumber \\
    C &\equiv \mathbb{K}_{ij}^{-+} = \frac{\omega}{2} \left[\Theta_{ij} + \left(\frac{\mu^2}{\omega^2}\right) \Omega_{ij} +  i \zeta \left(\frac{\mu}{\omega}\right) \epsilon_{ij} \right] e^{-\omega L} \nonumber \\
    D &\equiv \mathbb{K}_{ij}^{--} = \mathbb{K}_{ij}^{++} = A,
\end{align}
where we used $\omega \equiv \sqrt{\vert\bold{k}\vert^2 + \mu^2}$ here for notational ease. 

The objects above are written in terms of the tensors $\{\Theta_{ij}, \Omega_{ij}, \epsilon_{ij}\}$ which follow a simple algebra, simplifying the computation. In particular, we will use the helpful identities $\varepsilon \Theta\varepsilon = - \Omega$ and $\varepsilon \Omega\varepsilon = - \Theta$, as well as the standard properties of the projectors $\Theta$ and $\Omega$. Doing so, we get
\begin{align}
    A-BD^{-1}C &=\left[1 - e^{-2\omega L}(1-\zeta^2)\right] D.
\end{align}
To compute the determinant of these $2 \times 2$ matrices, it is convenient to use the well-known formula 
\begin{align}
    \det X_{2 \times 2} = \frac{1}{2}\left[\left({\rm tr} X\right)^2 - {\rm tr}\left(X^2\right)\right].
\end{align}
In the end, the determinant for the PMC case is given by
\begin{align} \label{DetFinalPMC}
    \det \mathbb{K}_{\rm PMC} =\frac{\mu^4}{16}\left[1 - e^{-2\omega L}(1-\zeta^2)\right]^2.
\end{align}
Before proceeding, let us discuss a bit this expression.

First, by taking the limit $\mu \to 0$ here, the determinant vanishes, differently from the PEC case. This feature comes from the photon mass, not from its CS origin, being present also in the Abelian limit of the CF model~\cite{Dudal26a}. This is due to the zero-mode that appears in the $\mu \to 0$ limit for PMC. Indeed, in this limit,  the operator $\mathbb{K}_{\rm PMC}$ becomes transverse, with an emergent boundary gauge symmetry $b_i^\gamma \to b_i^\gamma + \beta^\gamma k_i$ which must be fixed. However, since this factor $\mu^4$ is $L$-independent, it will not contribute to the Casimir energy. 

Furthermore, if we take the limit $\zeta \to 0$ while keeping $\mu \neq 0$, we throw away the information about the CS origin of the mass, but keep the massive character of the photon. Computing the Casimir energy in this limit, we get
\begin{align}
    \mathcal{E}_{\rm PMC}^{\zeta \to 0} &= \frac{1}{2} \int \! \frac{d^2 \bold{k}}{(2\pi)^2} 2 \log\left[1 - e^{-2 L \sqrt{\vert \bold{k}\vert^2 +\mu^2}}  \right] \nonumber \\
    &= 2 \, \mathcal{E}_{\rm PEC}
\end{align}
That is, we recover exactly the Casimir energy obtained in the Abelian limit of the CF model~\cite{Dudal26a}. In particular, we also obtain the mismatch between PEC and PMC cases. This raises a puzzle in the massless limit similar to the van Dam-Veltman-Zakharov discontinuity~\cite{vanDam70, Zakharov70} because for Maxwell theory the results of PEC and PMC should match. The resolution of this puzzle is related with the emergent boundary symmetry that we already mentioned before (for more details, see~\cite[Sec.~VI]{Dudal26a}).

Finally, we discuss a feature that stands out. Remember that the parameter $\zeta$ was included only to keep track of the CS term contribution, but in the MCS theory, we have $\zeta = 1$. In this case, we have $\det \mathbb{K}_{\rm PMC} =\frac{\mu^4}{16}$, which although non-zero, is $L$-independent. Thus, for PMC wires in a MCS electrodynamics there is no Casimir effect at all. This is a distinguishing feature of the CS term, since the result obtained through the Abelian limit of the CF model is different. Therefore, it shows that under suitable conditions, the Casimir effect (or its absence) can tell us about the CS origin of the photon mass.

%%%%%%%%%%%%%%%%%%%%%%%%%%%%%%%%%%%%%%%%%%%%%%%%%%%%%%
\subsection{PMC2 wires}

After the peculiar vanishing Casimir force in the PMC case, it is natural to wonder if by relaxing one of its constraints one can obtain a non-zero Casimir energy. Considering the PMC2 case, we have the operator
\begin{align}
\mathbb{K}_{\rm PMC2} = 
\begin{pmatrix}
    \mathbb{K}_{xx}^{++} & \mathbb{K}_{xx}^{+-} \\
    \mathbb{K}_{xx}^{-+} & \mathbb{K}_{xx}^{--} 
\end{pmatrix},
\end{align}
where the entries of this $2 \times 2$ matrix are given by Eq.\eqref{MCSKopPMC2}. Its determinant can be immediately computed:
\begin{align}
   \det  \mathbb{K}_{\rm PMC2} =  \frac{\left( k_t^2 + \mu^2\right)^2}{4 \left(\vert \bold{k} \vert^2 + \mu^2\right)} \left(1 - e^{-2 L \sqrt{\vert \bold{k}\vert^2 +\mu^2}} \right). 
\end{align}
One can readily notice that after discarding the $L$-independent factors that are irrelevant for the Casimir effect, we find exactly the same vacuum energy functional obtained imposing PEC conditions, that is
\begin{align}
    \mathcal{E}_{\rm PMC2} &= \frac{1}{2} \int \! \frac{d^2 \bold{k}}{(2\pi)^2} \log\left[1 - e^{-2 L \sqrt{\vert \bold{k}\vert^2 +\mu^2}}  \right] \nonumber \\ 
    &= \mathcal{E}_{\rm PEC}.
\end{align}
That is, for the MCS electrodynamics, the Casimir energy between two perfectly conducting parallel wires (PEC) is exactly the same as for two infinitely permeable wires (PMC2). This agrees with the results of~\cite{Alves10}. Moreover, since the parameter $\zeta$ is not present, the result for the Abelian CF model would be exactly the same. This result is completely different from the PMC case. 

After presenting these results, we remark that it would be a non-trivial task to guess them only by looking at the boundary dynamical operator. In fact, although $\mathbb{K}_{\rm PEC} \neq \mathbb{K}_{\rm PMC2}$ (even in the Maxwell limit), they yield the same Casimir energy. Moreover, $\mathbb{K}_{\rm PMC}$ has a more intricate tensorial structure, has an exponential factor similar to the other operators, but in the end there is an exact cancellation and the Casimir energy vanishes.

%%%%%%%%%%%%%%%%%%%%%%%%%%%%%%%%%%%%%%%%%%%%%%%%%%%%
\section{Intermezzo: Discussing the results} \label{SecIntermezzo}

Let us briefly discuss the results obtained so far and try to grasp their physical meaning. For two identical wires, the PEC-PEC and PMC2-PMC2 conditions give exactly the same Casimir energy, whereas the interaction vanishes for PMC-PMC. This contrast is particularly striking because the PMC case gives a non-zero Casimir force for both Maxwell theory and CF model. Moreover, the PMC2 condition obtained by relaxing one of the PMC constraints gives a non-zero force in the MCS theory.

A simple way to get a physical intuition about these results is the following. The MCS theory contains one physical degree of freedom, but the boundary requires solutions that simultaneously satisfy the equations of motion and the boundary conditions on the wire. The Casimir energy is known to be sensitive to the number of propagating degrees of freedom, which act as messengers, providing channels for the interaction between the wires. The counting of propagating degrees of freedom compatible with each boundary condition is done in App.~\ref{AppCountingDOF}.

This analysis provides a clear indication of the reason why we have a vanishing result for the PMC case. The covariant PMC condition imposes two constraints on the wires. For $\mu \neq 0$, the MCS equations couple these components to the remaining one, and only the trivial solution survives when PMC is imposed. Thus, the disappearance of the propagating mode is caused by the imposition of a too restrictive boundary condition in a theory in which the CS term relates the electric and magnetic fields.

In the Maxwell theory and the CF model, the term that forces the last component to vanish is absent, and a nontrivial mode survives under PMC conditions, showing that the cancellation is a consequence of the MCS equations. This observation is reinforced by our explicit computation of the Casimir energy since in the MCS theory there is a non-trivial cancellation of the $L$-dependent part between parity-even and parity-odd contributions. If the CS contribution is removed while the photon mass is retained, as in the CF model, that cancellation no longer occurs and the Casimir energy would be non-zero.

In the PEC and PMC2 cases, after imposing the constraints on the wires, we have one propagating degree of freedom. The boundary dynamical operators are different in each case, but their $L$-dependent part are the same. Since the $L$-independent pieces disappear after subtracting the vacuum energy of the free space, these boundary conditions yield the same Casimir energy. One might think that this equality could be explained by mode counting alone. However, as we will see in the following, the counting establishes that these configurations possess one active channel, but does not guarantee that they yield the same Casimir force.
 
Thus, we are led to consider other boundary conditions to check if the same pattern will persist. We would like to analyze two situations: first, a case in which we do not have any propagating degree of freedom to check if this will give a vanishing Casimir force again; second, a case in which we have only one propagating degree of freedom, to see if having only one active channel is enough to guarantee that we have the same Casimir energy as the PEC-PEC and PMC2-PMC2 cases. Thankfully, we have all the necessary ingredients in our hands, we only need to mix the already considered boundary conditions.

Consider first a PEC wire facing a covariant PMC wire. We show in App.~\ref{AppCountingDOF} that imposing these conditions on the wires in a MCS theory, only the trivial solution is allowed. Hence, the mode counting suggests that the PEC-PMC interaction must vanish. We will see that the presence of the PMC condition for one wire is already sufficient to kill the single MCS propagating mode, at least for the cases studied here.  This behavior is again specific to the MCS theory; in the Maxwell theory and in the Abelian limit of the CF model this mixed boundary condition provides a non-zero repulsive Casimir force.

The PEC-PMC2 configuration provides the complementary test. Since PMC2 does not impose the additional condition $E_y=0$, the MCS constraints allow one physical degree of freedom, and mode counting indicates that we have a non-zero Casimir energy. It does not, however, predict its value or sign. Indeed, the analysis done in App.~\ref{AppCountingDOF} already indicates that the surviving mode obeys constraints that are different from the ones previously found for two identical boundaries. As the explicit calculation will show, we will find a repulsive Casimir force rather than the attractive force found before in the PEC-PEC and PMC2-PMC2 cases (and with different absolute value also). Interestingly enough, in the Maxwell theory, we find the same result as we have found in the PEC-PMC case; for the Abelian limit of the CF model, the result is different from all the other cases.

These observations make clear both the value and the limitation of the counting argument. The existence of a boundary-compatible mode is a necessary indicator of a non-vanishing Casimir force for the configurations considered here, but the Casimir energy is not determined solely by the number of such modes. That is, mode counting only determines whether there is an active channel or not, whereas the corresponding magnitude and sign of the Casimir force will depend on other factors.

%%%%%%%%%%%%%%%%%%%%%%%%%%%%%%%%%%%%%%%%%%%%%%%%%%%%%%%%%%%%%%%%%%%%%%%%%%%%%%%%%%%%%%%%%%%%%%
\section{Mixed boundary conditions} \label{SecMixedBC}

\subsection{Mixing PEC and PMC}
We now investigate the case in which we impose PEC on one wire while applying PMC to the other, that is,
\begin{align} \label{MixedBoundCond}
   \tilde{F}_\mu n_\mu \big\vert_{\Sigma_-} &= 0, \nonumber \\
    F_{\mu \nu} n_\nu \big\vert_{\Sigma_+} &= 0.
\end{align}
The computation follows the same steps as before, {\it mutatis mutandis}, so we present the results more directly.

To impose the mixed boundary condition in the wires, we define $G = \tilde{F}_\mu n_\mu$ and $G_\mu = F_{\mu \nu} n_\nu$ as before, and introduce the following term in the action:
\begin{align}
      S_{\rm BC} = \int_{\Sigma^-} d^2 \bold{x} \, b (\bold{x}) G + \int_{\Sigma^+} d^2 \bold{x} \, b_i (\bold{x}) G_i,
\end{align}
where $b$ and $b_i$ are auxiliary fields acting as Lagrange multipliers to impose PEC and PMC boundary conditions on $\Sigma_-$ and $\Sigma_+$, respectively. We remark that PEC is implemented through a scalar condition $G\vert_{\Sigma_-} = 0$ while PMC is imposed using a vector condition $G_i\vert_{\Sigma_+} = 0$.

The terms imposing the boundary conditions can be rewritten in a manifestly Lorentz-covariant form~\eqref{Action_BC}, where the anti-symmetric tensor $H_{\mu \nu}$ is defined here as
\begin{align}
    H_{\mu \nu} = \delta(y - y_-) i \epsilon_{\mu \nu \rho} n_\rho b + \delta(y - y_+) \left(b_\mu n_\nu - b_\nu n_\mu\right). 
\end{align}
The next steps are the same as before: we write the full action and perform a suitable shift to decouple the photon. The partition function gets factorized and we focus in the boundary contribution since this is the relevant part to compute the Casimir energy after the shift. The effective boundary action is given formally by Eq~\eqref{BoundEffAct1}, which we repeat here for convenience
\begin{align}
    S_b = -\frac{1}{2} \! \int \!\!\frac{d^3k}{(2\pi)^3} \left[ k_\mu H_{\mu \rho}(k) \left(K^{-1}\right)_{\rho \sigma} k_\nu H_{\nu \sigma}(-k) \right].
\end{align}
For mixed boundary conditions we have, in Fourier space,
\begin{align}
    H_{\mu \nu}(k) =  e^{i k_y y_-} h_{\mu \nu}^-(\bold{k}) + e^{i k_y y_+} h_{\mu \nu}^+(\bold{k}) ,
\end{align}
where we defined the auxiliary tensors
\begin{align}
    h_{\mu \nu}^-(\bold{k}) &= i \epsilon_{\mu \nu \rho} n_\rho \, b(\bold{k}), \nonumber \\
    h_{\mu \nu}^+(\bold{k}) &= \left( \delta_{i \mu} n_\nu - \delta_{i \nu} n_\mu\right) b_i(\bold{k}).
\end{align}
We can rewrite the effective boundary action~\eqref{BoundEffAct1} as
\begin{align}
    S_b &= -\frac{1}{2} \! \int \!\!\frac{d^3k}{(2\pi)^3} \left(K^{-1}\right)_{\rho \sigma} k_\mu  k_\nu R_{\mu \rho \nu \sigma},
\end{align}
where the tensor $R_{\mu \rho \nu \sigma}$ is given by
\begin{align}
    R_{\mu \rho \nu \sigma}  &=  -\epsilon_{\mu \rho \alpha}    \epsilon_{\nu \sigma \beta} n_\alpha n_\beta b(\bold{k}) b(-\bold{k}) \nonumber \\
    &+ e^{i k_y (y_- - y_+)} i \epsilon_{\mu \rho \alpha} n_\alpha  \left( \delta_{j \nu} n_\sigma - \delta_{j \sigma} n_\nu\right)   b(\bold{k}) b_j(-\bold{k}) \nonumber \\
    &+ e^{i k_y (y_+ - y_-)} \left( \delta_{i \mu} n_\rho - \delta_{i \rho} n_\mu\right)  i \epsilon_{\nu \sigma \beta} n_\beta   b_i(\bold{k}) b(-\bold{k}) \nonumber \\
    &+  \! \left( \delta_{i \mu} n_\rho \!- \delta_{i \rho} n_\mu\right) \!\! \left( \delta_{j \nu} n_\sigma \!- \delta_{j \sigma} n_\nu\right) \! b_i(\bold{k}) b_j(-\bold{k}).
\end{align}
One can notice that the first and last lines of the expression above have the same tensor structure that we have found when studying the PEC and PMC cases previously. 

After some computational effort, we find
\begin{align}
    S_b &= -\frac{1}{2}\int \frac{d^3 k}{(2\pi)^3}\left[ b(\bold k) \left(\frac{-\bold k^2}{k^2 + \mu^2}\right)b(-\bold k)\right. \nonumber\\
    &\left. + b_i (\bold k) \left(\frac{ k^2 \delta_{ij} - \bold k^2 \Theta_{ij}(\bold k) + \mu k_y \epsilon_{ij}}{k^2 + \mu^2}\right)b_j (-\bold k) \right.\nonumber\\
    &\left. + i b(\bold k) e^{-i k_y L}\left( \frac{\zeta\mu k_j - k_y k_i\epsilon_{ij} }{k^2 + \mu^2}\right) b_j (-\bold k)\right. \nonumber\\
    &\left. + i b_i (\bold k) e^{i k_y L} \left( \frac{-\zeta\mu k_i + k_y \epsilon_{ij} k_j}{k^2 + \mu^2}\right) b(-\bold k)\right].
\end{align}
Now, performing the integral in $dk_y$, we obtain
\begin{align}
    S_b &= \frac{1}{2}\int \frac{d^2 \bold k}{(2\pi)^2}\left[ b(\bold k) \left(\frac{\bold k^2}{ 2 \omega}\right)b(-\bold k)\right. \nonumber\\
    &\left. + b_i (\bold k) \left( \frac{\omega}{2}\Theta_{ij} + \frac{\mu^2}{2\omega} \Omega_{ij}\right)b_j (-\bold k) \right.\nonumber\\
    &\left. + i b(\bold k) e^{-\omega L}\left( - \zeta\frac{\mu}{2\omega} k_j - \frac{i}{2}k_i \epsilon_{ij} \right) b_j (-\bold k)\right. \nonumber\\
    &\left. + i b_i (\bold k) e^{-\omega L} \left( \zeta\frac{\mu}{2\omega} k_i - \frac{i}{2} \epsilon_{ij} k_j\right) b(-\bold k)\right],
\end{align}
where we defined, as before, $\omega \equiv \sqrt{\vert\bold{k}\vert^2 + \mu^2}$. The last expression can be rewritten in a more convenient form:
\begin{align} \label{SbMixed}
S_b \! &= \!\frac{1}{2}\!\int \!\!\! \frac{d^2 \bold k}{(2\pi)^2} \! \begin{pmatrix}
    b(\bold{k}) & b_i(\bold{k}) 
\end{pmatrix} \!
\begin{pmatrix}
    \mathbb{K}^{++} & \mathbb{K}^{+-}_j \\
    \mathbb{K}^{-+}_i & \mathbb{K}^{--}_{ij} 
\end{pmatrix} \!
\begin{pmatrix}
      b(-\bold{k}) \\
      b_j(-\bold{k}) 
\end{pmatrix}, 
\end{align}
where each block in the above expression is given by
\begin{align}
    \mathbb{K}^{++} &= \frac{\bold k ^2}{2\omega}, \nonumber \\
    \mathbb{K}_{ij}^{--} &= \frac{\omega}{2} \Theta_{ij} + \frac{\mu^2}{2\omega} \Omega_{ij},  \nonumber \\
    \mathbb{K}_{j}^{+-} &= i\left[-\zeta\frac{\mu}{2\omega} k_j - \frac{i}{2} k_\ell \epsilon_{\ell j}  \right]e^{-\omega L}, \nonumber \\
    \mathbb{K}^{-+}_i &= i\left[\zeta\frac{\mu}{2\omega} k_i - \frac{i}{2} \epsilon_{i\ell} k_\ell \right]e^{-\omega L} .
\end{align}
The next step is to write the vacuum energy using the partition function and perform the Gaussian functional integral in the boundary fields. To compute the determinant of the boundary dynamical operator we use Eq.~\eqref{detblock}. We already know that
\begin{align}
    \det \mathbb{K}^{--}_{ij} = \frac{\mu^2}{4}.
\end{align}
To compute the other determinant, we need to first get
\begin{align}
    W \equiv \mathbb{K}^{++} - \mathbb{K}^{+-}_i \left(\mathbb{K}^{--}\right)^{-1}_{ij} \mathbb{K}^{-+}_j.
\end{align}
For the second term in the above expression, we find
\begin{align}
    \mathbb{K}^{+-}_i \left(\mathbb{K}^{--}\right)^{-1}_{ij} \mathbb{K}^{-+}_j &= -e^{-2\omega L}\left[ - \zeta^2 \frac{1}{2\omega}\bold k^2 \right. \nonumber \\
    &\left. + \frac{1}{4}(\epsilon k)_i \left(\frac{2}{\omega}\Theta_{ij} + \frac{2\omega}{\mu^2}\Omega_{ij}\right)(\epsilon k)_j\right].
\end{align}
It will be useful to use the intermediate expressions
\begin{align}
    (\epsilon k)_i \Theta_{ij} (\epsilon k)_j &= \vert\bold{k}\vert^2, \nonumber \\
    (\epsilon k)_i \Omega_{ij} (\epsilon k)_j &= 0.
\end{align}
Therefore, we find for the relevant determinant: 
\begin{align}
    \det  \mathbb{K}^{\rm Mix1} &=  \frac{\mu^2 \vert\bold{k}\vert^2}{8\omega}\left[1 + e^{-2\omega L}(1-\zeta^2)\right].
\end{align}

Some comments are in order. First, note that by taking the limit $\mu \to 0$ at this point we find a vanishing determinant, analogously to the PMC case. Once again, this is due to the presence of a zero-mode in the boundary dynamical operator in this limit, as it can be seen by the purely transverse structure of the $\mathbb{K}^{--}_{ij}$ block in the limit $\mu \to 0$. Remember that in the determinant computation done here, the inverse of such block was calculated in an intermediate step. Since this factor does not depend on $L$, it will not contribute to the Casimir energy in the end. 

Now, taking the limit $\zeta \to 0$, we throw away the information peculiar to the CS term, while keeping the photon massive. Then, discarding the $L$-independent terms that are irrelevant here, the Casimir energy reads
\begin{align} \label{VacEnergyMixedBC}
    \mathcal{E}_{\rm Mix1}^{\zeta \to 0} &= \frac{1}{2} \int \frac{d^2 \bold{k}}{(2\pi)^2} \log \left[1 + e^{-2 L \sqrt{\vert\bold{k}\vert^2 + \mu^2}}\right] \nonumber \\
    %&= \frac{1}{4\pi} \int_0^{\infty} \! dk \, k  \log\left[1 + e^{-2 L \sqrt{k^2 +\mu^2}}  \right] \nonumber \\
    &=  \frac{1}{4\pi} \int_\mu^{\infty} \! dz \, z  \log\left[1 + e^{-2 L z}  \right] \nonumber \\
    %&= \frac{1}{48\pi L^2} \left[8 L^3 \mu^3 + 6 L \mu \, {\rm Li}_2\left(-e^{2L\mu}\right) - 3 {\rm Li}_3\left(-e^{2L\mu}\right)\right] \nonumber \\
    &=\! -\frac{1}{16\pi L^2} \left[ 2 L \mu \, {\rm Li}_2\left(-e^{-2L\mu}\!\right) \!+\!  {\rm Li}_3\left(-e^{-2L\mu}\!\right)\right]\!.
    \end{align}
This would be the result for the Casimir energy with a CF-like massive photon (without a topological origin) when we consider two parallel wires under the mixed boundary conditions described in Eq.~\eqref{MixedBoundCond}. The result is a repulsive Casimir force for any finite value of $\mu$. 

Furthermore, let us see what happens in the Maxwell limit. Taking the limit $\mu \to 0$ in Eq.~\eqref{VacEnergyMixedBC}, we obtain:
\begin{align}
    \mathcal{E}^{\mu \to 0}_{\rm Mix1} = \frac{3 \zeta(3)}{64\pi L^2}.
\end{align}
Note that the sign is the opposite to what we have obtained using PEC plates in the limit $\mu \to 0$, indicating a repulsive force instead of an attractive one. Moreover,
\begin{align}
    \mathcal{E}^{\mu \to 0}_{\rm Mix1} = -\frac{3}{4} \lim_{\mu \to 0} \mathcal{E}^{\rm PEC}.
\end{align}

Finally, considering $\zeta = 1$ --- the current case of interest --- we see that there is no Casimir effect anymore, like in the case with two wires under PMC conditions. Therefore, for MCS theory, imposing PMC on one wire seems to be enough to annihilate the Casimir effect. This could have been expected from the analysis done in App.~\ref{AppCountingDOF}, reinforcing the interpretation relating the Casimir energy with the mode counting. 

It is important to emphasize the difference between our results and that of~\cite{Alves10}. There, the authors considered mixed boundary conditions consisting of an infinitely permeable wire and a perfectly conducting wire in MCS electrodynamics, obtaining a non-vanishing and repulsive Casimir force for the case $\mu \neq 0$. There is no tension between our results and theirs because the PMC condition adopted here is not equivalent to having an infinitely permeable wire. We will consider this scenario below, extracting the relevant information from the previous computation, finding perfect agreement between our results in the PEC-PMC2 scenario with those of~\cite{Alves10}. 

%%%%%%%%%%%%%%%%%%%%%%%%%%%%%%%%%%%%%%%%%%%%%%%%%%%
\subsection{Mixing PEC and PMC2}

Finally, we analyze the case PEC-PMC2. Following the same strategy as before, we can use the PMC result and impose $b_t =0$ to effectively annihilate the constraint $E_y=0$. In doing so, we get an operator of the form
\begin{align}
\mathbb{K}_{\rm Mix2} = 
\begin{pmatrix}
    \mathbb{K}^{++} & \mathbb{K}_{x}^{+-} \\
    \mathbb{K}_{x}^{-+} & \mathbb{K}_{xx}^{--} 
\end{pmatrix},
\end{align}
where each component here is given by
\begin{align}
    \mathbb{K}^{++} &= \frac{\bold k ^2}{2 \omega}, \nonumber \\
    \mathbb{K}_{xx}^{--} &= \frac{\left(k_t^2 + \mu^2\right)}{2 \omega}  \nonumber \\
    \mathbb{K}_{x}^{+-} &= \left[-\zeta\frac{i\mu}{2\omega} k_x + \frac{1}{2} k_t \right]e^{-\omega L} \nonumber \\
    \mathbb{K}^{-+}_x &= \left[\zeta\frac{i\mu}{2\omega} k_x - \frac{1}{2} k_t \right]e^{-\omega L}.
\end{align}
Thus, the determinant reads 
\begin{align}
    \det \!\mathbb{K}_{\rm Mix2} \!=\! \frac{1}{4 \omega^2} \!\left[\! \vert\bold{k}\vert^2 \! \left(k_t^2 \!+\! \mu^2\right) \!+\! e^{-2 \omega L} \!\left(\omega k_t \!-\! \zeta i \mu k_x\right)^2 \!\right].
\end{align}
Let us analyze some limits before proceeding. First of all, by taking the Maxwell limit $\mu \to 0$, we obtain
\begin{align}
    \det \mathbb{K}^{\mu \to 0}_{\rm Mix2} = \frac{k_t^2}{4} \left[1 + e^{-2 \vert \bold{k} \vert L}\right].
\end{align}
Then, using this expression, we find the Casimir energy, 
\begin{align}
        \mathcal{E}^{\mu \to 0}_{\rm Mix2} &= \frac{1}{2} \int \frac{d^2 \bold{k}}{(2\pi)^2} \log \left(1 + e^{-2 \vert \bold{k} \vert L} \right) \nonumber \\
        &= \frac{3 \zeta(3)}{64 \pi L^2} =  \mathcal{E}^{\mu \to 0}_{\rm Mix1},
\end{align}
exactly as we have found before. Thus, we conclude that for Maxwell electrodynamics, having parallel wires under PEC-PMC or under PEC-PMC2 gives the same result. 

Now, we consider the limit $\zeta \to 0$, which turns off the CS term while keeping the mass. In this case, we have
\begin{align}
    \det \mathbb{K}^{\zeta \to 0}_{\rm Mix2} \!=\! \frac{1}{4 \omega^2} \left[\vert\bold{k}\vert^2 \! \left(k_t^2 \!+\! \mu^2\right) \!+ e^{-2 \omega L}\!  \left(\vert\bold{k}\vert^2 \!+\! \mu^2\right) k_t^2 \right]. %\nonumber \\
   % &= \frac{\bold{k}^2 \! \left(k_t^2+\mu^2\right)}{4 \omega^2} \left[1 + e^{-2 \omega L} \frac{\left(\bold{k}^2 + \mu^2\right) k_t^2}{\bold{k}^2 \! \left(k_t^2+\mu^2\right)} \right].
\end{align}
The above result leads to the following Casimir  energy:  
\begin{align}\label{NotComplexIntegralMixedBC}
        \mathcal{E}^{\zeta \to 0}_{\rm Mix2} &= \frac{1}{2} \int \frac{d^2 \bold{k}}{(2\pi)^2} \log \left[1 + e^{-2 \omega L} \frac{\left(\vert\bold{k}\vert^2 + \mu^2\right) k_t^2}{\vert\bold{k}\vert^2 \! \left(k_t^2+\mu^2\right)} \right].
\end{align}
Using polar coordinates, we can write: $k_t = \vert \bold{k} \vert \cos\theta$ and $k_x = \vert \bold{k} \vert \sin\theta$. We know that $\omega^2 = \vert \bold{k} \vert^2 + \mu^2 = k_t^2 + k_x^2 + \mu^2$. Thus, the argument of the logarithm reads
\begin{align}
    1 \!+\! e^{-2 \omega L} \frac{\left(\! \vert\bold{k}\vert^2 \!+\! \mu^2 \! \right) \! k_t^2}{\vert\bold{k}\vert^2 \! \left(k_t^2 \!+\! \mu^2\right)} \! &= \! 1 + e^{-2 \omega L} \frac{\omega^2 \cos^2\theta}{\vert \bold{k} \vert^2 \cos^2\theta + \mu^2} \nonumber \\
    &=\! \frac{\left(\vert \bold{k} \vert^2 + e^{-2\omega L} \omega^2\right) \cos^2\theta + \mu^2}{\vert \bold{k} \vert^2 \cos^2\theta + \mu^2}.
\end{align}
Hence the logarithm splits into a difference of two terms that have the common form $\log\left[a \cos^2\theta + b\right]$. Since $d^2 \bold{k} = \vert \bold{k} \vert d\vert \bold{k} \vert d\theta = \omega d\omega d\theta$ (remember that $\omega^2 = \vert \bold{k} \vert^2 + \mu^2$), the integral to be done can be rewritten in the form
\begin{align}
     \mathcal{E}^{(\zeta \to 0)}_{\rm Mix2} = \frac{1}{8\pi^2} \int \! \omega d\omega d\theta &\bigg\lbrace \log\left[a \cos^2\theta + b\right] \nonumber \\
     &- \log\left[a' \cos^2\theta + b\right] \bigg\rbrace,
\end{align}
where $a \equiv \omega^2\left(1 + e^{-2\omega L}\right) - \mu^2$, $a' \equiv \left(\omega^2 - \mu^2\right)$, and $b \equiv \mu^2$. Now, if $A \geq 0$ and $B > 0$, we can use the following  result,
\begin{align} \label{IntLogEq}
    \int_{0}^{2\pi} \!\!\!\! d\theta  \log\left[A \cos^2\theta + B\right] \!=\! 4 \pi \log\left(\frac{\sqrt{B} +\! \sqrt{A+B}}{2}\right), 
\end{align}
which can be obtained through the integral w.r.t.~$A$ of 
\begin{align} \label{IntLogEq2}
    \int_{0}^{2\pi} \!\!\!\! d\theta  \frac{1}{A \cos^2\theta + B} & = 2\int_{0}^{2\pi} \!\!\!\! dx  \frac{1}{A/2 \cos x + B+A/2}.
 \end{align}
with as initial condition that $\eqref{IntLogEq}= 2\pi \log B$ at $A=0$, next to using the standard substitution for the kind of integral \eqref{IntLogEq2}, viz.~$t=\tan (x/2)$.

Using the above expression \eqref{IntLogEq}, we obtain
\begin{align}
     \mathcal{E}^{(\zeta \to 0)}_{\rm Mix2} = \frac{1}{2\pi} \int_\mu^{\infty} \! d\omega \, \omega \log\left[\frac{\mu + \omega \sqrt{1 + e^{-2 \omega L}}}{\mu + \omega}\right].
\end{align}
This is the result for the Casimir energy between two parallel wires under the mixed PEC-PMC2 obtained from the Abelian limit of the CF model. The corresponding force can be obtained readily from $f=-\frac{\partial \mathcal{E}}{\partial L}$, giving us
\begin{align} \label{CasForceMix2Zeta0}
    f^{\zeta \to 0}_{\rm Mix2} = \frac{1}{2\pi} \int_\mu^{\infty} \! d\omega \, \frac{\omega^3}{\left[\omega + e^{2 \omega L}\left(\omega + \mu \sqrt{1 + e^{-2 \omega L}} \right)\right]}. 
\end{align}

Finally, we consider the case $\zeta=1$. Notice that in this case, the squared modulus of the complex number $Z \equiv \left(\omega k_t - i \mu k_x\right) = \vert Z \vert  e^{i\phi}$ is exactly the first term appearing in the above expression, that is,
\begin{align}
    \vert \omega k_t - i \mu k_x \vert^2  &= \vert\bold{k}\vert^2 \left(k_t^2 + \mu^2\right), 
\end{align}
Thus, using polar coordinates, we can conveniently rewrite $Z^2 = \vert\bold{k}\vert^2 \left(k_t^2 + \mu^2\right) e^{2 i \phi}$. Doing so, the determinant expression can be rewritten in a simpler form as
\begin{align}
     \det \mathbb{K}_{\rm Mix2}^{\zeta \to 1} = \frac{\vert\bold{k}\vert^2 \left(k_t^2 + \mu^2\right)}{4 \omega^2}\left[1 + e^{-2 \omega L} e^{2 i \phi}\right].
\end{align}
Putting this result in the vacuum energy expression and discarding $L$-independent terms as usual, we obtain
\begin{align} \label{ComplexIntegralMixedBC}
    \mathcal{E}_{\rm Mix2}^{\zeta \to 1} = \frac{1}{2} \int \frac{d^2 \bold{k}}{(2\pi)^2} \log \left(1 + e^{-2 \omega L} e^{2 i \phi}\right).
\end{align}
We can conveniently rewrite this integral as
\begin{align} 
    \mathcal{E}_{\rm Mix2}^{\zeta \to 1} &= \frac{1}{2} \int \frac{d^2 \bold{k}}{(2\pi)^2}\bigg\lbrace\frac12 \log \left(1 + e^{-2 \omega L} e^{2 i \phi}\right) \nonumber \\
    &+ \frac12 \log \left(1 + e^{-2 \omega L} e^{-2 i \phi}\right)\bigg\rbrace \nonumber \\
    &= \frac{1}{4} \int \frac{d^2 \bold{k}}{(2\pi)^2} \log \big\vert 1 + e^{-2 \omega L} e^{2 i \phi}\big\vert^2,
\end{align}
where in the second term we made $k_x \to -k_x$, which changed $\phi \to -\phi$, allowing us to rewrite the integral in a manifestly real manner. Denoting $\chi \equiv e^{-2\omega L}$, we have 
\begin{align}
    \mathcal{E}_{\rm Mix2}^{\zeta \to 1} &= \frac{1}{4\pi} \int_\mu^\infty d\omega \, \omega \, I(\omega), 
\end{align}
where we used polar coordinates with $k_t = \vert \bold{k} \vert \cos\theta$ and $k_x = \vert \bold{k} \vert \sin\theta$, changed variables $\vert\bold{k}\vert d\bold{k} = \omega d\omega$, and defined 
\begin{align}
     I(\omega) \equiv \frac{1}{4\pi} \int_0^{2\pi} \! d\theta \log\left[1 + \chi^2 + 2\chi \cos(2 \phi)\right],
\end{align}
Now, using $\cos(2 \phi) = 1 -2\sin^2\phi$, we can write
\begin{align}
    \cos(2 \phi) &= 1- 2\frac{\mu^2 k_x^2}{\vert\bold{k}\vert^2 \left(k_t^2 + \mu^2\right)} \nonumber \\
    &= 1 - 2 \frac{\mu^2 \sin^2\theta}{\vert\bold{k}\vert^2 \cos^2\theta + \mu^2}.
\end{align}
Thus, we can rewrite the logarithm above, obtaining
\begin{align}
   I &= \!\frac{1}{4\pi}\!\! \int \!\!\! d\theta \log\bigg\lbrace\frac{\left[(1+\chi)^2 \vert\bold{k}\vert^2 + 4\chi\mu^2\right]\cos^2\!\theta + (1-\chi)^2 \mu^2}{\vert\bold{k}\vert^2 \cos^2\theta + \mu^2}\bigg\rbrace.   
   \end{align}
The integral above can be written as the difference of two expressions of the form $\log\left[A \cos^2\theta + B\right]$. As before, we can use Eq.~\eqref{IntLogEq} to solve these angular integrals. The result after simplification can be written as 
   \begin{align}
   I =  \log\left[1 + \chi\left(\frac{\omega - \mu}{\omega + \mu}\right)\right].
\end{align}
Therefore, the final result for the energy is
\begin{align}
    \mathcal{E}_{\rm Mix2}^{\zeta \to 1} = \frac{1}{4\pi} \int_\mu^\infty \! d\omega \, \omega \log\left[1 + e^{-2  \omega L}\left(\frac{\omega - \mu}{\omega+\mu}\right)\right].
\end{align}
From the above expression, we can immediately compute $f=-\frac{\partial \mathcal{E}}{\partial L}$, finding for the Casimir force the expression
\begin{align} \label{CasForceMix2Zeta1}
    f_{\rm Mix2}^{\zeta \to 1} = \frac{1}{2\pi} \int_\mu^\infty \! d\omega \,  \frac{ \omega^2 (\omega - \mu)}{(\omega - \mu) + e^{2 \omega L} (\omega + \mu)}.
\end{align}
This is in perfect agreement with the expression reported in~\cite{Alves10}, where the authors computed the Casimir effect in a MCS electrodynamics between a perfectly conducting wire and an infinitely permeable one using the energy-momentum tensor and Green's functions techniques.

Interestingly enough, comparing the results obtained above for the Casimir force in the limits $\zeta \to 0$ (Abelian CF) and $\zeta \to 1$ (MCS), one can analytically show (see App.~\eqref{AppIneqMixedZeta}) that the latter is smaller than the former for any $L>0$ and $\mu>0$. That is, for PEC-PMC2 we have: 
\begin{align}
    f_{\rm Mix2}^{\zeta \to 1}\left(L,\mu\right) < f_{\rm Mix2}^{\zeta \to 0}\left(L,\mu\right) < \frac{3 \zeta(3)}{32 \pi L^3}, \quad \forall L, \mu >0.
\end{align}
We conclude that for PEC-PMC2 conditions the Casimir force is more suppressed when the photon mass comes from a CS term than when it comes from a CF term. Moreover, in both cases they give a smaller value than the Maxwell result ($\mu = 0$) showing that the presence of a mass for the photon weakens the Casimir force in comparison with the massless limit, as intuitively expected.

%%%%%%%%%%%%%%%%%%%%%%%%%%%%%%%%%%%%%%%%%%%%%%%%%%%%%%%%%%%%%%%%%%%%%%%%%%%%%%%%%%%%%%%%%%%%%%%%%%%%%%%%%%

\begin{table*}[t]
\caption{\label{TableSummary}%
Casimir energy per unit wire length for the boundary conditions considered here, where $L$ is the distance between the wires and $\mu>0$ is the photon mass. Negative (positive) energies correspond to attractive (repulsive) interactions.}
\begin{ruledtabular}
\begin{tabular}{lccc}
Boundary conditions
& Maxwell\footnote{Notice that this column refers to the Maxwell result, which could be different from the $\mu \to 0$ limit of a MCS theory in some cases.} ($\mu = 0$)
& CF ($\zeta = 0$, $\mu>0$)
& MCS ($\zeta = 1$, $\mu>0$)
\\
\colrule
PEC-PEC
& $-\zeta(3)/16\pi L^2$
& $\mathcal E_{-}$~\cite{Dudal26a}
& $\mathcal E_{-}$~\cite{Milton90}
\\[0.6ex]
PMC-PMC
& $-\zeta(3)/16\pi L^2$
& $2\mathcal E_{-}$~\cite{Dudal26a}
& $0$
\\[0.6ex]
PMC2-PMC2
& $-\zeta(3)/16\pi L^2$
& $\mathcal E_{-}$
& $\mathcal E_{-}$~\cite{Alves10}
\\[0.6ex]
PEC-PMC
& $+3 \zeta(3)/64\pi L^2$
& $\mathcal E_{+}$
& $0$
\\[0.6ex]
PEC-PMC2
& $+3 \zeta(3)/64\pi L^2$
& $\mathcal E_{\mathrm{CF}}^{\mathrm{mix}}$
& $\mathcal E_{\mathrm{MCS}}^{\mathrm{mix}}$~\cite{Alves10}
\end{tabular}
\end{ruledtabular}

\begin{minipage}{1.0\textwidth}

\begin{align}
\begin{cases}
\mathcal E_{-} &\equiv -\frac{1}{16\pi L^{2}}\left[ 2 L \mu \,\mathrm{Li}_{2}\left(e^{-2 L \mu}\right)
+\mathrm{Li}_{3}\left(e^{-2 L \mu}\right)\right], \vspace{0.2cm}\\
\mathcal E_{+} &\equiv -\frac{1}{16\pi L^{2}}\left[
2 L\mu \,\mathrm{Li}_{2}\left(-e^{-2 L \mu}\right)
+\mathrm{Li}_{3}\left(-e^{-2 L \mu}\right)\right], \vspace{0.2cm}\\
\mathcal E_{\mathrm{CF}}^{\mathrm{mix}} &\equiv \frac{1}{2\pi} \int_{\mu}^{\infty}d\omega\,\omega
\log\left[ \frac{ \mu+\omega\sqrt{1+e^{-2\omega L}}}{\mu+\omega} \right], \vspace{0.2cm}\\
\mathcal E_{\mathrm{MCS}}^{\mathrm{mix}} &\equiv \frac{1}{4\pi} \int_{\mu}^{\infty}d\omega\,\omega
\log\left[ 1+e^{-2\omega L} \left(\frac{\omega-\mu}{\omega+\mu}\right) \right].
\end{cases}
\end{align}

\end{minipage}
\end{table*}

%%%%%%%%%%%%%%%%%%%%%%%%%%%%%%%%%%%%%%%%%%%%%%%%%%%%%%%%%%%%%%%%%%%%%%%%%%%%%%%%%%%%%%%%%%%%%%
\section{Conclusion} \label{SecConclusion}

In this work, we investigated the Casimir effect between two parallel wires in MCS electrodynamics in $(2+1)D$ for several phenomenologically motivated boundary conditions. Using a functional approach in which the boundary conditions are implemented through auxiliary fields localized on the interfaces, we derived an effective boundary theory whose functional determinant allows us to compute the Casimir energy of the system. This framework allowed us to treat PEC, PMC, and PMC2 conditions, as well as the mixed PEC-PMC and PEC-PMC2 cases, within a unifying formalism. The final results for the Casimir energies are gathered in Table~\ref{TableSummary}.

For two perfectly conducting wires, we recovered the standard MCS Casimir energy originally obtained by other methods~\cite{Milton90}. This result coincides with the one obtained from the CF model~\cite{Dudal26a}, thus the PEC Casimir effect can detect the photon mass, but cannot distinguish if it comes from a CS term. The same conclusion holds for two infinitely permeable wires, the PMC2 result is identical to the PEC one~\cite{Alves10}. In both cases, the CS contribution leaves no fingerprint on the Casimir energy, other than the photon mass, and the Maxwell result is smoothly recovered when $\mu \to 0$.

For the PMC condition, the situation is rather different. In this case, the CS term contribution enters the boundary dynamical operator and exactly cancels the $L$-dependent factor in the functional determinant. Consequently, there is no more Casimir effect between two wires under PMC conditions in a MCS theory. Interestingly enough, this cancellation is not obtained in the CF model~\cite{Dudal26a}, indicating that the vanishing Casimir force in this case is genuinely tied to the CS origin of the mass. Therefore, this shows that, under suitable boundary conditions, the Casimir effect (or its absence) can probe not only the photon mass, but also distinguish its CS origin.

For all the reflection-symmetric geometries considered in this work (PEC-PEC, PMC-PMC, PMC2-PMC2), we found no example of a repulsive Casimir force, though it vanishes in one case. This suggests that the mere presence of the parity-breaking CS term is not sufficient by itself to produce a loophole in the no-go theorem~\cite{Kenneth06}.

For PEC-PMC boundaries, the MCS Casimir force again vanishes, indicating that imposing the PMC condition in one wire can be already enough to kill off the Casimir effect. Once again, the CF model result is non-vanishing for this case, providing another example where the Casimir effect can distinguish the CS origin of the photon mass. By contrast, the PEC-PMC2 case presents a non-vanishing Casimir force. Our result for this case perfectly matches the already known result for a perfectly conducting wire facing an infinitely permeable one~\cite{Alves10}. Moreover, for the PEC-PMC2 case, we found an inequality showing that the MCS theory result is smaller than the CF model one for any $L, \mu >0$. In the Maxwell limit, the PMC and PMC2 cases studied here become indistinguishable as far as the Casimir effect is concerned. For $\mu \neq 0$, however, their distinction becomes physical. 

There is an appealing interpretation in terms of the propagating degrees of freedom compatible with the boundaries that could allow us to guess some results obtained here. The configurations for which a propagating MCS mode survives (PEC-PEC, PMC2-PMC2, and PEC-PMC2) present a non-vanishing Casimir force, whereas the PMC condition eliminates the propagating mode, consequently killing the Casimir force. This provides a simple physical intuition for the otherwise rather striking cancellations found in the explicit computation. On the other hand, the comparison between the cases with an active channel shows that counting degrees of freedom is not enough to fully determine the Casimir force details. In our understanding, these results are ``perhaps not surprising, but hardly obvious"~\cite{Milton90}.

A promising avenue for future work is to explore the interplay between our functional framework and the Symanzik approach for local QFTs with boundaries~\cite{Symanzik81}, in which one consider boundary conditions emerging from first-principles constructions rather than imposing them phenomenologically~\cite{Blasi10, Bertolini26, Maggiore18}. In particular, we intend to explore interconnections between our approach and the very recent work~\cite{Maggiore26}, where the authors analyze a system closely related to the one considered here. Furthermore, it would be interesting to compute the Casimir energy in the parity-invariant version of the MCS theory~\cite{Hagen92}, to check which features described here would change in a scenario with topological mass and parity-symmetry. 

Finally, we close this paper with an intriguing question: would it possible to adapt the proof of the no-go theorem~\cite{Kenneth06} for a MCS electrodynamics, in which the bulk dynamics exhibits an inherent parity violation?

%%%%%%%%%%%%%%%%%%%%%%%%%%%%%%%%%%%%%%%%%%%%%%%%%%%%%%%%%%%%%%%%%%%%%%%%%%%%%%%%%%%%%%%%%%%%%%
\begin{acknowledgments}

P.~De~Fabritiis is grateful to Sebbe Stouten for the interesting discussions and acknowledges the National Council for Scientific and Technological Development - CNPq for the financial support (No. 402459/2024-5). The work of D.~Dudal was supported by KU Leuven IF project C14/21/087. 

\end{acknowledgments}
%%%%%%%%%%%%%%%%%%%%%%%%%%%%%%%%%%%%%%%%%%%%%%%%%%%%%%%%%%%%%%%%%%%%%%%%%%%%%%%%%%%%%%%%%%%%%%%%%%%%%%%%%%%%%%%%%%%%%%%%%%%%%%%%%%%%%%%%%%%%%%%%
\appendix

\section{Counting degrees of freedom} \label{AppCountingDOF}

In this Appendix, we will briefly discuss the number of (classically) propagating degrees of freedom in MCS electrodynamics searching for some plane-wave solutions. Then, we will impose different boundary conditions on the wires to see how many degrees of freedom will survive in each case. For convenience, we will perform this analysis adopting the Minkowskian signature, with $\eta_{\mu \nu} = \textrm{diag}(+, -, -)$, but the conclusions are similar for its Euclidean counterpart. We will adopt $\epsilon^{012} = \epsilon_{012} = +1$. Let us consider the Minkowskian $2+1$ MCS action,
\begin{align}
    S = \int d^3 x \left(-\frac{1}{4}F^{\mu\nu}F_{\mu\nu} + \frac{\mu}{2}\epsilon^{\mu\rho\nu}A_\mu \partial_\rho A_\nu\right).
\end{align}
The equation of motion following from this action is
\begin{align}
    \partial_\nu F^{\nu\mu} + \frac{\mu}{2}\epsilon^{\mu\nu\rho}F_{\nu\rho}=0
\end{align}
Defining $\tilde{F}^\mu = \frac{1}{2}\epsilon^{\mu\nu\rho}F_{\nu\rho}$, we immediately find that 
\begin{align} \label{FtilEq1}
    \partial_\mu \tilde{F}^\mu = 0.
\end{align}
Using $\tilde{F_\mu}$, we can rewrite the equation of motion as
\begin{align} \label{FtilEq2}
    \epsilon^{\mu\nu\rho}\partial_\nu \tilde{F}_\rho - \mu \tilde{F}^\mu = 0.
\end{align}
Now, contracting the above equation with $\epsilon_{\beta\sigma\mu}\partial^\sigma$ and using the property $\partial_\mu \tilde{F}^\mu = 0$, one can get the condition:
\begin{align} \label{FtilEq3}
    (\partial_\mu \partial^\mu + \mu^2)\tilde{F}_\beta = 0.
\end{align}
Thus, $\tilde{F}_\mu$ is subject to three constraints:~\eqref{FtilEq1},\eqref{FtilEq2},\eqref{FtilEq3}. 

Let us propose the following {\it ansatz} for this analysis:
\begin{align} \label{MCSansatz}
\tilde{F}_\beta (t,x,y) = G_\beta(y) e^{i\omega t + ik_x x}. 
\end{align}
Plugging this {\it ansatz} in Eq.~\eqref{FtilEq3}, we obtain
\begin{align}
    G''_\beta +k_y^2 G_\beta = 0,
\end{align}
where $G'_\beta = \frac{dG_\beta}{dy}$ and $k_y^2 = \omega^2 - k_x^2 - \mu^2$. The solution for the differential equation above can be written as
\begin{align} \label{SolEOMansatz}
    G_\beta(y) = A_\beta \sin\left(k_y y\right) + B_\beta  \cos\left(k_y y\right),
\end{align}
where the coefficients $A_\beta$ and $B_\beta$ are complex numbers. 

Considering now the constraint of Eq.~\eqref{FtilEq2}, we find
\begin{align} \label{AnsatzCond1}
    ik_x G_2 - G'_1 - \mu G_0 &= 0 \nonumber \\
    - i\omega G_2 + G'_0 + \mu G_1 &= 0 \nonumber \\
     i\omega G_1 - ik_x G_0 + \mu G_2 &= 0.
\end{align}
Finally, imposing Eq.~\eqref{FtilEq1}, we get
\begin{align} \label{AnsatzCond2}
i\omega G_0 - ik_x G_1 - G_2' = 0.
\end{align}
Thus, for our {\it ansatz}~\eqref{MCSansatz}, the general solution of the equations of motion~\eqref{SolEOMansatz} is subject to the constraints~\eqref{AnsatzCond1} and~\eqref{AnsatzCond2}. In the following, we will impose the PEC and PMC boundary conditions and see how many degrees of freedom are left in the end.

%%%%%%%%%%%%%%%%%%%%%%%%%%%%%%%%%%%%%%%%%%%%%%%%
\subsection{PEC boundary condition}

For PEC boundary conditions, we have 
\begin{align}
    \tilde{F}_\mu n^\mu \big\vert_{\Sigma_\pm}= 0.
\end{align}
Since we have $n_\mu = (0, 0, 1)$ and $\Sigma_\pm$ is the subspace defined by $y = \pm L/2$, in practice, we have $\tilde{F}_2 \left(y = \pm L/2\right) = 0$, which for our {\it ansatz} reads
\begin{align}
    G_2 \left(\pm L/2\right) = 0.
\end{align}
Considering our general solution~\eqref{SolEOMansatz}, these equations give us two possibilities: i) $B_2 = 0$ and $k_y = \frac{2\pi}{L} n$, with $n \in \mathbb{Z}$; ii) $A_2 = 0$ and $k_y = \frac{2\pi}{L} \left(n+\frac12\right)$, with $n \in \mathbb{Z}$. We choose the first option (this does not change the conclusion), finding
\begin{align}
    G_2 = A_2 \sin(k_y \, y), \quad {\rm with} \quad k_y = \frac{2\pi n}{L}.
\end{align}
Now, we check the consequence of PEC condition on Eq.~\eqref{AnsatzCond1}. On the wires ($y = \pm L/2$), we have
\begin{align} \label{EqConstPart1}
    \left(G'_1 + \mu G_0\right) \big\vert_{y = \pm L/2} &= 0 \nonumber \\
     \left(G'_0 + \mu G_1\right) \big\vert_{y = \pm L/2} &= 0 \nonumber \\
     \left(\omega G_1 - k_x G_0\right) \big\vert_{y = \pm L/2} &= 0.
\end{align}
Since $k_y = \frac{2\pi n}{L}$, we already know that $\sin\left(k_y \frac{L}{2}\right) = 0$ and $\cos\left(k_y \frac{L}{2}\right) = \pm 1$. Thus, from conditions~\eqref{EqConstPart1} we get
\begin{align}
     k_y A_1 + \mu B_0 &= 0\nonumber \\
     k_y A_0 + \mu B_1 &= 0 \nonumber \\
     \omega B_1 - k_x B_0  &= 0
\end{align}
Then, we find for these coefficients
\begin{align}
    B_1 &= \frac{k_x}{\omega} B_0 \nonumber \\
    A_1 &= -\frac{\mu}{k_y} B_0 \nonumber \\
    A_0 & = -\frac{\mu k_x}{k_y \omega} B_0.
\end{align}
Thus, so far, we have already obtained 
\begin{align}
    G_0(y) &= B_0 \left[-\frac{\mu k_x}{\omega k_y}\sin (k_y y) + \cos (k_y y)\right]\nonumber \\
     G_1(y) &= B_0 \left[-\frac{\mu }{ k_y}\sin (k_y y) + \frac{k_x}{\omega}\cos (k_y y)\right] \nonumber \\
     G_2(y) &= A_2 \sin (k_y y).
\end{align}
We still need to impose Eq.~\eqref{AnsatzCond2} at $y = \pm L/2$. We get,
\begin{align}
  i \omega B_0 \left(1 - \frac{k_x^2}{\omega^2}\right) - A_2 k_y = 0.
\end{align}
That is, we can also write $B_0$ in terms of $A_2$. This means that the only free coefficient after all is $A_2$ which is complex, having two degrees of freedom. Finally, we need to impose a reality condition in $\tilde{F}_\mu$ since its components are the electric and magnetic fields. 

From \eqref{MCSansatz}, the general solution for $\tilde{F}_\mu$ admits the following Fourier expansion:
\begin{align}
    \tilde{F}_\mu(x) = \frac{1}{(2\pi)^2}\int d\omega ~dk_x G_\mu(\omega,k_x;y) e^{i(\omega t + k_x x)}
\end{align}
The reality condition $\tilde{F}^*_\mu = \tilde{F}_\mu $ will give us $G^*_\mu(\omega,k_x;y) = G_\mu(-\omega,-k_x;y)$, and since all the coefficients are proportional to $A_2$, we must have $A^*_2 (\omega,k_x) = A_2(-\omega,-k_x)$.

In summary, after imposing the PEC boundary conditions and the bulk equations of motion, all coefficients are determined by a single complex amplitude $A_2$. The reality condition relates the amplitude of the $(\omega,k_x)$ mode to that of its complex conjugate $(-\omega,-k_x)$ mode. Thus the conjugate Fourier modes are not independent, and only one physical polarization propagates.

Furthermore, note that considering the case $\mu = 0$ would not change this feature in the PEC case, we would still have one propagating degree of freedom here.

For completeness, we should also mention that there is another class of potential solutions. These correspond to surface modes with $G_2\equiv0$, $G_1=\pm G_0$ for $\omega=\pm k_x$, in which case $G_0\sim e^{\mp \mu y}$. As our $y$ extends over the full real line, no normalizable solution exists though.  However, even if $y$ was restricted to live on the interval $[-L/2,L/2]$, these surface modes would be irrelevant for the Casimir effect since their dispersion relation does not depend on $L$.  Also, for other combinations discussed below, such surface modes can appear as special solutions, but we will always discard those for similar reasons as just given.

%%%%%%%%%%%%%%%%%%%%%%%%%%%%%%%%%%%%%%%%%%%%%%
\subsection{PMC boundary condition}

For PMC boundary conditions, we have
\begin{align}
    F^{\mu \nu} n_\nu \big\vert_{\Sigma_\pm} = 0
\end{align}
Again, we have $n_\mu = (0, 0, 1)$ and $\Sigma_\pm$ is defined by $y = \pm L/2$. We can rewrite $F^{\mu\nu} = \epsilon^{\mu\nu\rho}\tilde{F}_\rho$. Thus, we get $\tilde{F}_0 \left(y = \pm L/2\right) = \tilde{F}_1 \left(y = \pm L/2\right) = 0$. Using the same reasoning as before, we find that 
\begin{align}
    G_0(y) &= A_0 \sin (k_y y),  \nonumber \\
    G_1(y) &= A_1 \sin (k_y y),
\end{align}
where $ k_y = \frac{2\pi n}{L}$ with $n \in \mathbb{Z}$. Now, looking at the third line of Eq.~\eqref{AnsatzCond1}, we see that having $G_0 \big\vert_{\Sigma_{\pm}} = G_1 \big\vert_{\Sigma_{\pm}} = 0$ imply that $G_2 \big\vert_{\Sigma_{\pm}} = 0$ also. This means that we have $G_2(y) = A_2 \sin (k_y y)$. As a consequence, using the first two lines of Eq.~\eqref{AnsatzCond1}, we find that $G'_0\big\vert_{\Sigma_{\pm}} = G'_1\big\vert_{\Sigma_{\pm}} = 0$. Furthermore, using Eq.~\eqref{AnsatzCond2}, we also see that $G'_2\big\vert_{\Sigma_{\pm}} = 0$. 

By the discussion  above, since at the wires we must have $G_0 = G_1 = G_2 = G'_0 = G'_1 = G'_2 = 0$, the only possible solution is the trivial one, with $A_0 = A_1 = A_2 = 0$. Therefore, we see that there are no propagating degrees of freedom for the MCS electrodynamics under the covariant PMC boundary conditions. 

Furthermore, if we take the limit $\mu \to 0$ in the beginning, having $G_0 \big\vert_{\Sigma_{\pm}} = G_1 \big\vert_{\Sigma_{\pm}} = 0$ would not imply that $G_2 \big\vert_{\Sigma_{\pm}} = 0$ anymore, since the term with $G_2$ in the last line of Eq.~\eqref{AnsatzCond1} would not be there if $\mu \to 0$. Therefore, in the limit $\mu = 0$, we would be able to find a propagating degree of freedom even under PMC boundary conditions.

%%%%%%%%%%%%%%%%%%%%%%%%%%%%%%%%%%%%%%%%%%%%%%%%%%%
\subsection{PMC2 boundary condition}

In the case of PMC2, we have the boundary conditions
\begin{align}
    F_{1 \nu} n_\nu \big\vert_{\Sigma_\pm} = 0.
\end{align}
Remembering that we have the relation $F^{\mu\nu} = \epsilon^{\mu\nu\rho}\tilde{F}_\rho$, this reads $\tilde{F}_0 \left(y = \pm L/2\right)=0$. Thus, 
\begin{align} 
    G_0 \left(\pm \frac{L}{2}\right) = 0.
\end{align}
Thus, we have
\begin{align}
G_0(y) = A_0 \sin k_y y,     
\end{align}
where $k_y = \frac{2n\pi}{L}$ with $n \in \mathbb{Z}$. The other components are
\begin{align}
    G_1(y) &= A_1 \sin(k_y y)  + B_1 \cos(k_y y) \nonumber \\
     G_2(y) &= A_2 \sin(k_y y) +   B_2 \cos(k_y y).
\end{align}
Considering that we have $G_0 \left(\pm \frac{L}{2}\right) = 0$, Eq.~\eqref{AnsatzCond1} gives us 
\begin{align} 
    ik_x G_2 - G'_1 \big\vert_{y = \pm L/2} &= 0, \nonumber \\
    - i\omega G_2 + G'_0 + \mu G_1 \big\vert_{y = \pm L/2} &= 0, \nonumber \\
     i\omega G_1 + \mu G_2 \big\vert_{y = \pm L/2} &= 0.
\end{align}
From the equations above, we get
\begin{align}
      A_1  &= \frac{ik_x}{k_y} B_2, \nonumber \\
   B_2  &= \frac{ik_y\omega}{\mu^2 -\omega^2} A_0 , \nonumber \\
     B_1  &= - \frac{\mu}{i \omega} B_2.
\end{align}
Now, imposing Eq.~\eqref{AnsatzCond2}, we find
\begin{align}
    i k_x B_1 + k_y A_2 = 0
\end{align}
\begin{align}
    A_2 =   \frac{ k_x\mu}{ k_y \omega} B_2.
\end{align}
We can see that after imposing all the constraints, we can determine all the coefficients in terms of $A_0$. As already done before, we still need to impose a reality condition on it. Therefore, in the end, we have one degree of freedom.

%%%%%%%%%%%%%%%%%%%%%%%%%%%%%%%%%%%%%%%%%%%%%%%%
\subsection{Mixed boundary condition 1}

%From the discussion done above, we conclude that the MCS electrodynamics has one propagating degree of freedom if we impose PEC  boundary conditions in the wires and none if we impose PMC conditions on them. 
Now, we consider the case in which we impose PEC in one wire and PMC in the other. That is, we consider: 
\begin{align} \label{MixedBC}
   \tilde{F}_\mu n_\mu \big\vert_{\Sigma_-} &= 0, \nonumber \\
    F_{\mu \nu} n_\nu \big\vert_{\Sigma_+} &= 0.
\end{align}
In practice, for our {\it ansatz}, this gives us the conditions
\begin{align}
    G_2 \left(-L/2\right) &= 0, \nonumber \\
    G_0 \left(+L/2\right) &= 0, \nonumber \\
    G_1 \left(+L/2\right) &= 0.
\end{align}
First, we consider the constraints~\eqref{AnsatzCond1} for $y=-L/2$:
\begin{align} \label{AppAuxEq3}
      \left(G'_1 + \mu G_0 \right)\big\vert_{y = - L/2} &= 0 \nonumber \\
     \left(G'_0 + \mu G_1\right) \big\vert_{y = - L/2} &= 0 \nonumber \\
     \left(\omega G_1 - k_x G_0\right) \big\vert_{y = - L/2} &= 0.
\end{align}
On the other hand, 
\begin{align} \label{AppAuxEq1}
    \left(ik_x G_2 - G'_1 \right)\big\vert_{y = + L/2} &= 0 \nonumber \\
    \left(- i\omega G_2 + G'_0 \right)\big\vert_{y = + L/2} &= 0 \nonumber \\
      \left(\mu G_2\right)\big\vert_{y = + L/2} &= 0.
\end{align}
In particular, we note that $G_2\left(\pm L/2\right) = 0$. Thus, as we already discussed before, we can write
\begin{align}
    G_2(y) = A_2 \sin(k_y \, y), \quad {\rm with} \quad k_y = \frac{2\pi n}{L}.
\end{align}
For $G_0$ and $G_1$ we write the more general expression
\begin{align}
    G_0(y) = A_0 \sin (k_y y)  + B_0 \cos (k_y y),\nonumber \\
     G_1(y) = A_1 \sin (k_y y) +   B_1 \cos (k_y y).
\end{align}
Furthermore, using $G_2\left(+L/2\right)= 0$ in the first two lines of Eq.~\eqref{AppAuxEq1} we find 
\begin{align} \label{AppAuxEq2}
    G'_1 \left(+L/2\right) &= 0 \nonumber \\
    G'_0 \left(+L/2\right) &= 0.
\end{align}
Now, due to the constraint $k_y = \frac{2\pi n}{L}$, we will have at the boundaries $\left(y=\pm L/2\right)$ the sine functions vanishing and the cosine functions giving $\pm 1$. Thus, the equations above~\eqref{AppAuxEq2} will imply that $A_0 = A_1 = 0$. Then, since the sine functions vanish in $y=-L/2$, we will have $ G'_0 \left(-L/2\right) =  G'_1 \left(-L/2\right) = 0$ also. However, plugging this information into Eq.~\eqref{AppAuxEq3}, since $\mu \neq 0$, we find  $G_0 \left(-L/2\right) =  G_1 \left(-L/2\right) = 0$. This is only possible if $B_0 = B_1 = 0$. Together with $A_0=A_1=0$, this means that 
\begin{align}
    G_0 = G_1 = 0.
\end{align}

Up to now, the only non-trivial profile is $A_2$. However, we still have to impose Eq.~\eqref{AnsatzCond2}. Considering that we have $G_0=G_1=0$, this implies that $G_2'=0$. But we already know that $G_2(y) = A_2 \sin(\frac{2\pi n}{L}\, y)$, thus, $A_2 = 0$.

Therefore, we conclude that also under mixed boundary conditions~\eqref{MixedBC}, in which we impose PEC in one wire and PMC in the other, there are no propagating degrees of freedom for the MCS electrodynamics.

%%%%%%%%%%%%%%%%%%%%%%%%%%%%%%%%%%%%%%%%%%%%%%%%
\subsection{Mixed boundary condition 2}

Finally, we will analyze the last case, where we have PEC in one wire and PMC2 in the other. The analysis in this case is more involved. Here we have:
\begin{align} \label{MixedBC}
   \tilde{F}_\mu n_\mu \big\vert_{\Sigma_-} &= 0, \nonumber \\
    F_{1 \nu} n_\nu \big\vert_{\Sigma_+} &= 0.
\end{align} 
The first condition gives us $\tilde{F}_2 \left(-L/2\right) = 0$. The second one gives $\tilde{F}_0 \left(+ L/2\right)$. Before imposing these conditions, let us discuss the constraints that come from Eqs.~\eqref{AnsatzCond1} and~\eqref{AnsatzCond2} for a general $y$ using Eq.~\eqref{SolEOMansatz}. From the sines,
\begin{align}
  i k_x A_2 + k_y B_1 - \mu A_0  &= 0, \nonumber \\
  -i \omega A_2 - k_y B_0 + \mu A_1  &= 0, \nonumber \\
   i \omega A_1 - i k_x A_0 + \mu A_2 &= 0, \nonumber \\
   i \omega A_0 - i k_x A_1 + k_y B_2 &=    0.
\end{align}
On the other hand, from the cosines,
\begin{align}
   i k_x B_2 - k_y A_1 - \mu B_0    &=   0,  \nonumber \\
   -i \omega B_2 + k_y A_0 + \mu B_1 &=  0,  \nonumber \\
   i \omega B_1 - i k_x B_0 + \mu B_2 &=  0,  \nonumber \\
    i \omega B_0 - i k_x B_1 - k_y A_2 &=  0.
\end{align}
From the above equations, we can obtain $A_0$ and $B_0$:
\begin{align}
    A_0 &= \frac{\omega A_1 -i \mu A_2}{k_x}\nonumber \\
    B_0 &= \frac{\omega B_1 -i \mu B_2}{k_x}.
\end{align}
The next step is to find $A_1$ and $A_2$. Using the relation $\omega^2 = k_x^2 + k_y^2 +\mu^2$, we find
\begin{align} 
\begin{pmatrix}
i (k_y^2+ \mu^2) & \mu \omega \\
   -\mu \omega & i (k_x^2+ \mu^2)  
\end{pmatrix} 
\begin{pmatrix}
    A_1 & \\
    A_2 & 
\end{pmatrix} = \begin{pmatrix}
    -k_x k_y B_2 & \\
    -k_x k_y B_1 & 
\end{pmatrix}.
\end{align}
Solving this system, we get
\begin{align}
    A_1 &= \frac{-\mu \omega B_1 + i (k_x^2 + \mu^2) B_2}{k_x k_y}, \nonumber \\
    A_2 &= \frac{\mu \omega B_2 + i (k_y^2 + \mu^2) B_1}{k_x k_y}.
\end{align}
Thus, substituting back in $A_0$, we find
\begin{align}
    A_0 = \frac{-\mu B_1 + i \omega B_2}{k_y}.
\end{align}
Thus, at this point, we have all the coefficients written in terms of $B_1$ and $B_2$. The equations that were not used are automatically satisfied once we plug in the solutions. 

Now, we will impose the relevant boundary conditions. In terms of $G_\beta = A_\beta \sin\left(k_y y\right) + B_\beta  \cos\left(k_y y\right)$, we have
\begin{align}
    G_2 \left(-L/2\right) &= 0, \nonumber \\
    G_0 \left(+L/2\right) &= 0.
\end{align}
Defining $\gamma \equiv  k_y L/2$ and the more economic notation $s_\gamma \equiv \sin(\gamma); c_\gamma \equiv \cos(\gamma)$, this gives us
\begin{align}
    -A_2 s_\gamma  + B_2 c_\gamma &= 0, \nonumber \\
    A_0 s_\gamma + B_0 c_\gamma &= 0.
\end{align}
These can be written in terms of $B_1$ and $B_2$ as
\begin{align}
 -i (k_y^2 + \mu^2) s_\gamma B_1 + \left[k_x k_y c_\gamma -\mu \omega s_\gamma \right] B_2    &= 0, \nonumber \\
\left[\omega k_y c_\gamma -\mu k_x s_\gamma\right]  B_1 + i\left[\omega k_x  s_\gamma -\mu k_y c_\gamma \right] B_2    &= 0.
\end{align}
To find a non-trivial solution for this homogeneous system, we need a vanishing determinant, which upon the use of trigonometric identities can be written as
\begin{align}
    k_x k_y \left[\mu k_x \sin(2 \gamma) -\omega k_y \cos(2\gamma)\right] = 0.
\end{align}
Considering $k_x, k_y \neq 0$, we find the condition:
\begin{align}
    \tan\left(k_y L\right) = \frac{\omega k_y}{\mu k_x}.
\end{align}
This equation gives a quantization condition for $k_y$ (reasonably more complicated than in the previous cases). At a $k_y$ satisfying this constraint, the $2 \times 2$ homogeneous system for $(B_1, B_2)$ has rank 1. Thus, we have 1 complex degree of freedom. Imposing the usual reality condition, we conclude that the mixed PEC-PMC2 boundary condition allows only one degree of freedom.

%%%%%%%%%%%%%%%%%%%%%%%%%%%%%%%%%%%%%%%%%%%%%%%%%%%%%%%%%%%%%%%%%%%%%%
\section{Inequality relating Casimir forces in PEC-PMC2 for $\zeta=0$ and $\zeta=1$}\label{AppIneqMixedZeta}

The goal of this Appendix is to analytically show that:
\begin{align}
    f_{\rm Mix2}^{\zeta \to 1}\left(L,\mu\right) < f_{\rm Mix2}^{\zeta \to 0}\left(L,\mu\right) < \frac{3 \zeta(3)}{32 \pi L^3}, \quad \forall L, \mu >0.
\end{align}

First, we can rewrite the Casimir forces of interest as 
\begin{align} \label{ForceTwoCasesI}
    f_I = \frac{1}{2\pi} \int_\mu^\infty \! d\omega \, \omega^2 e^{-2 \omega L} \varphi_I,
\end{align}
where $I=0$ represents the case $\zeta=0$~\eqref{CasForceMix2Zeta0} and $I=1$ the case $\zeta=1$~\eqref{CasForceMix2Zeta1}, with $\varphi_I$ given by
\begin{align}
    \varphi_0 &= 1/s \left(s + \tau\right),\nonumber \\
    \varphi_1 &= (1 - \tau) / [s^2 \left(1 - \tau\right) + 2 \tau], 
\end{align}
with $\tau$ and $s$ given by
\begin{align}
    \tau &\equiv \mu / \omega, \nonumber \\
    s &\equiv  \sqrt{1+ e^{-2 \omega L}}.
\end{align}
Now, we compute the difference $\delta\varphi \equiv \varphi_0 - \varphi_1$, obtaining
\begin{align}
    \delta\varphi = \frac{\tau \left[2 - s(1-\tau)\right]}{s(s+\tau)\left[s^2(1-\tau) + 2\tau\right]}.
\end{align}
We need to discuss the ranges of the variables. Since $e^{-2 \omega L} \in (0,1)$, we have $s = \sqrt{1 + e^{-2 \omega L}} \in (1, \sqrt{2})$. Since $\omega^2 = \bold{k}^2 + \mu^2$, we have $\tau = \mu /\omega \in (0,1]$, considering $\mu >0$.

We have $s \in (1,\sqrt{2})$ and $\tau \in (0,1]$, then we have $s(1-\tau)<\sqrt{2}<2$. Thus, the numerator of $\delta \varphi$ is strictly positive for $\mu>0$; the same for the denominator. Therefore, this proves that $\varphi_0 > \varphi_1$. Using Eq.~\eqref{ForceTwoCasesI}, this implies that we have $f_{\rm Mix2}^{\zeta \to 1}\left(L,\mu\right) < f_{\rm Mix2}^{\zeta \to 0}\left(L,\mu\right), \, \forall L, \mu >0$.

The next step is to show that the Maxwell result constitutes an upper bound. In the limit $\mu \to 0$, we have $\tau \to 0$. The immediate consequence is $\varphi_0, \varphi_1 \to 1/s^2$, the common Maxwell limit. We will now show that both $\varphi_0\left(L,\mu\right)$ and $\varphi_1\left(L,\mu\right)$ are smaller than $1/s^2$, $\forall L,\mu >0$.

First, $\varphi_0$ is obviously smaller than  $1/s^2$ since $s\tau >0$ for $L,\mu>0$. To prove that $\varphi_1 < 1/s^2$, since $1-\tau >0$, we only need to show that $2\tau > s^2 \tau$. But this follows immediately since $\tau \in (0,1]$ and $s \in (1,\sqrt{2})$. Therefore, it is proven.

%%%%%%%%%%%%%%%%%%%%%%%%%%%%%%%%%%%%%%%%%%%%%%%%%%%%%
%%%%%%%%%%%%%%%%%%%%%%%%%%%%%%%%%%%%%%%%%%%%%%%%%%%%%
	
%%%%%%%%%%%%%%%%%%%%%%%%%%%%%%%%%%%%%%%%%%%%%%%%%%%%%%%%%%%%%%%%%%%%%%%%%%%%%%%%%%%%%%%%%%%%%%%%%%%%%%%%%%%%%%%%%%%%%%%%%%%%%%%%%%%%%%%%%%%%%%%%%%%%%%%%%%%%%%%%%%%%%%%%%%%%%%%%%%%%%%%%%%%%%%%%%%%%%%%%%%%%%%%%%%%%%%%%%%%%%%%%%%%%%%%%%%%


\begin{thebibliography}{99}


    \bibitem{Casimir48}
    H. B. G. Casimir, {\it On the attraction between two perfectly conducting plates}, Proc. Kon. Ned. Akad. Wet. 51, 793 (1948).

    \bibitem{Plunien86}
    G. Plunien, B. Muller, and W. Greiner, {\it The Casimir effect}, Phys. Rep. 134, 87 (1986).

    \bibitem{Bordag01}
    M. Bordag, U. Mohideen, and V. M. Mostepanenko, {\it New developments in the Casimir effect}, Phys. Rep. 353, 1 (2001).

    \bibitem{Milton01}
    K. A. Milton, {\it The Casimir Effect: Physical Manifestations of Zero-Point Energy}, (World Scientific, Singapore, 2001) ISBN:9789810243975.

    \bibitem{Milton04}
    K. A. Milton, {\it The Casimir effect: Recent controversies and progress}, J. Phys. A 37, R209 (2004).

    \bibitem{Bordag09a}
    M. Bordag, G. Klimchitskaya, U. Mohideen, and V. Mostepanenko, {\it Advances in the Casimir effect}, Int. Ser. Monogr. Phys. 145, 1 (2009).

    \bibitem{Lamoreaux97}
    S. K. Lamoreaux, {\it Demonstration of the Casimir Force in the $0.6$ to $6 \mu m$ Range}, Phys. Rev. Lett. 78, 5 (1997).

    \bibitem{Bimonte21}
    G. Bimonte, B. Spreng, P. A. Maia Neto, G. L. Ingold, G. L. Klimchitskaya, V. Mostepanenko, and R. S. Decca, {\it Measurement of the Casimir Force between $0.2$ and $8 \mu m$: Experimental Procedures and Comparison with Theory}, Universe 7, 93 (2021).

    \bibitem{Deser82a}
    S. Deser, R. Jackiw, and S. Templeton, {\it Topologically massive gauge theories}, Ann. Phys. 140, 372 (1982).

    \bibitem{Deser82b}
    S. Deser, R. Jackiw, and S. Templeton, {\it Three-Dimensional Massive Gauge Theories}, Phys. Rev. Lett. 48, 975 (1982).

    \bibitem{Dunne99}
    G. V. Dunne, {\it Aspects of Chern-Simons theory}, arXiv:hep-th/9902115. 

    \bibitem{Fradkin13}
    E. Fradkin, {\it Field Theories of Condensed Matter Physics}, Cambridge University Press, 2013.

    \bibitem{Witten16}
    E. Witten, {\it Three lectures on topological phases of matter}, Riv. Nuovo Cimento 39, 313 (2016).

    \bibitem{Tong16}
    D. Tong, {\it Lectures on the Quantum Hall Effect}, arXiv:1606.06687.

    \bibitem{Witten26}
    E. Witten, {\it The Chern-Simons Function and the Quantum Hall Effect}, arXiv:2609.21182.

    \bibitem{Zhang89}
    S. C. Zhang, T. H. Hansson, and S. Kivelson, {\it Effective-Field-Theory Model for the Fractional Quantum Hall Effect}, Phys. Rev. Lett. 62, 82 (1989).

    \bibitem{Wilczek82a}
    F. Wilczek, {\it Magnetic Flux, Angular Momentum, and Statistics}, Phys. Rev. Lett. 48, 1144 (1982).

    \bibitem{Wilczek82b}
    F. Wilczek, {\it Quantum Mechanics of Fractional-Spin Particles}, Phys. Rev. Lett. 49, 957 (1982).

    \bibitem{Laughlin88}
    R. B. Laughlin, {\it The Relationship between High-Temperature Superconductivity and the Fractional Quantum Hall Effect}, Science 242, 525 (1988).

    \bibitem{Milton90}
    K. A. Milton and Y. J. Ng, {\it Maxwell-Chern-Simons Casimir effect}, Phys. Rev. D 42, 2875 (1990).

    \bibitem{Ambjorn83}
    J. Ambjørn and S. Wolfram, {\it Properties of the vacuum. I. Mechanical and thermodynamic}, Ann. Phys. (N.Y.) 147, 1 (1983).

    \bibitem{Milton92}
    K. A. Milton and Y. J. Ng, {\it Maxwell-Chern-Simons Casimir effect. II. Circular boundary conditions}, Phys. Rev. D 46, 842 (1992).
    
     \bibitem{Alves10}
    D. T. Alves, E. R. Granhen, J. F. Medeiros Neto, and S. Perez, {\it Repulsive Maxwell-Chern-Simons Casimir effect}, Phys. Lett. A 374, 2113 (2010). 

    \bibitem{Boyer74}
    T. H. Boyer, {\it Van der Waals forces and zero-point energy for dielectric and permeable materials}, Phys. Rev. A 9, 2078 (1974).

    \bibitem{Kenneth06}
    O. Kenneth and I. Klich, {\it Opposites Attract: A Theorem about the Casimir Force}, Phys. Rev. Lett. 97, 160401 (2006).


    \bibitem{Bachas07}
    C. P. Bachas, {\it Comment on the sign of the Casimir force}, J. Phys. A: Math. Theor. 40 9089 (2007).

    \bibitem{Asorey13}
    M. Asorey and J. M. Muñoz-Castañeda, {\it Attractive and repulsive Casimir vacuum energy with general boundary conditions}, Nucl. Phys. B 874, 852 (2013).
    
    \bibitem{Jiang19}
    Q.-D. Jiang and F. Wilczek, {\it Chiral Casimir forces: Repulsive, enhanced, tunable}, Phys. Rev. B 99, 125403 (2019).

    \bibitem{Grushin11}
    A. G. Grushin and A. Cortijo, {\it Tunable Casimir repulsion with three-dimensional topological insulators}, Phys. Rev. Lett. 106, 020403 (2011).

    \bibitem{Wilson15}
    J. H. Wilson, A. A. Allocca, and V. Galitski, {\it Repulsive Casimir force between Weyl semimetals}, Phys. Rev. B 91, 235115 (2015).
    
    \bibitem{Bordag00}
    M. Bordag and D. V. Vassilevich, {\it Casimir force between Chern-Simons surfaces}, Phys. Lett. A 268,  75 (2000).

    \bibitem{Bordag09b}
    M. Bordag, I. V. Fialkovsky, D. M. Gitman, and D. V. Vassilevich, {\it Casimir interaction between a perfect conductor and graphene described by the Dirac model}, Phys. Rev. B 80, 245406 (2009). 

    \bibitem{Marachevsky23}
    V. N. Marachevsky and A. A. Sidelnikov, {\it Casimir-Polder interaction with Chern-Simons boundary layers}, Phys. Rev. D 107, 105019 (2023).

    \bibitem{Fialkovsky18}
    I. Fialkovsky, N. Khusnutdinov, and D. Vassilevich, {\it Quest for Casimir repulsion between Chern-Simons surfaces}, Phys. Rev. B 97, 165432 (2018).

    \bibitem{Vassilevich20}
    D. Vassilevich, {\it On the (im)possibility of Casimir repulsion between Chern-Simons surfaces}, Mod. Phys. Lett. A 35, 2040017 (2020).

%%%%%%%%%%%%%%%%%%%%%%%%%%%%%%%%%%%%%%%%%%%%%%%%%%%
     \bibitem{Dudal26a}
    D. Dudal, P. De Fabritiis, and S. Stouten, {\it Non-Abelian Casimir energy in the Curci-Ferrari model through a functional approach},  J. High Energ. Phys. 2026, 87 (2026).

    \bibitem{Mandelstam76}
    S. Mandelstam, {\it Vortices and Quark Confinement in NonAbelian Gauge Theories}, Phys. Rept. 23, 245 (1976).

    \bibitem{Nambu74}
    Y. Nambu, {\it Strings, monopoles, and gauge fields}, Phys. Rev. D 10, 4262 (1974).

    \bibitem{tHooft81}
    G. ’t Hooft, {\it Topology of the Gauge Condition and New Confinement Phases in NonAbelian Gauge Theories}, Nucl. Phys. B 190, 455 (1981).

%%%%%%%%%%%%%%%%%%%%%%%%%%%%%%%%%%%%%%%%%%%%%%%%%%%
   
    
    \bibitem{Dudal20}
    D. Dudal, P. Pais, and L. Rosa, {\it Casimir energy in terms of boundary quantum field theory: The QED case}, Phys. Rev. D  102, 016026 (2020).

    \bibitem{Dudal22}
    F. Canfora, D. Dudal,  T. Oosthuyse, P. Pais, and L. Rosa, {\it The Casimir effect in chiral media using path integral techniques}, J. High Energ. Phys. {\bf 2022}, 95 (2022).
		
    \bibitem{Dudal23}
    T. Oosthuyse and  D. Dudal, {\it Interplay between chiral media and perfect electromagnetic conductor plates: Repulsive vs. attractive Casimir force transitions}, SciPost Phys. {\bf 15}, 213 (2023).

    \bibitem{Dudal24}
    D. Dudal, A. Gobeyn, T. Oosthuyse, S. Stouten, D. Vercauteren, {\it Casimir energy with perfect electromagnetic boundary conditions and duality: A field-theoretic approach}, Phys. Rev. D 110, 065015 (2024).


   \bibitem{Canfora25}
    F. Canfora, D. Dudal, T. Oosthuyse, L. Rosa, and S. Stouten, {\it Dynamical edge modes in Maxwell theory from a BRST perspective, with an application to the Casimir energy}, J. High Energ. Phys. 2025, 161 (2025).

    \bibitem{Dudal26b}
    J. Devroe, D. Dudal, and S. Stouten, {\it On the Casimir effect with mixed dynamical edge mode and perfect electromagnetic conducting boundary conditions}, arXiv:  2605.10662.

    \bibitem{Dudal26d}
    D. Dudal, P. De Fabritiis, S. Stouten, and D. Vercauteren, {\it The Casimir effect in Gribov-Zwanziger theory}, arXiv:2609.11448.

   
    \bibitem{Bordag85}
    M. Bordag, D. Robaschik, and E. Wieczorek, {\it Quantum Field Theoretic Treatment of the Casimir Effect}, Ann. Phys. 165, 192 (1985).

    
    \bibitem{Karabali25}
    D. Karabali, A. Maj, and V. P. Nair, {\it Casimir effect in Yang-Mills theories in three and four dimensions}, Phys. Rev. D 112, 094037 (2025).

    \bibitem{Karabali18}
    D. Karabali, V. P. Nair, {\it Casimir effect in (2 + 1)-dimensional Yang-Mills theory as a probe of the magnetic mass}, Phys. Rev. D 98, 105009 (2018).

    \bibitem{Karabali96}
    D. Karabali, V. P. Nair, {\it A gauge-invariant Hamiltonian analysis for non-Abelian gauge theories in (2+1) dimensions}, Nucl. Phys. B 464, 135 (1996).
    
    \bibitem{Symanzik81}
    K. Symanzik, {\it Schrödinger representation and Casimir effect in renormalizable quantum field theory}, Nucl. Phys. B 190, 1 (1981).

    \bibitem{Blasi10}
    A. Blasi, N. Maggiore, N. Magnoli, and S. Storace, {\it Maxwell–Chern–Simons theory with a boundary}, Class. Quantum Grav. 27, 165018 (2010).

    \bibitem{Maggiore18}
    N. Maggiore, {\it Holographic reduction of Maxwell-Chern-Simons theory}, Eur. Phys. J. Plus 133, 281 (2018).
    
    \bibitem{Bertolini26}
    E. Bertolini, M. Doyle, N. Maggiore, C. Murphy, and C. Piras, {\it Edge modes in Chern-Simons theory on a strip}, Phys. Rev. D 113, 085022 (2026).

     \bibitem{Maggiore26}
    N. Maggiore, {\it Edge physics and the Casimir interaction in Maxwell-Chern-Simons theory on a strip}, Phys. Rev. D 114, 045030 (2026).
    
%%%%%%%%%%%%%%%%%%%%%%%%%%%%%%%%%%%%%%%%%%%%%%%%%%%%%
    \bibitem{Hagen92}
    C. R. Hagen, {\it Parity conservation in Chern-Simons theories and the anyon interpretation}, Phys. Rev. Lett. 68,  3821 (1992).

    \bibitem{Wilczek92}
    F. Wilczek, {\it Disassembling anyons}, Phys. Rev. Lett. 69, 132 (1992).

    \bibitem{DelCima15}
    O. M. Del Cima, {\it The parity-preserving massive QED$_3$: vanishing $\beta$-function and no parity anomaly}, Phys. Lett. B 750, 1 (2015).

    \bibitem{Well20}
    W. B. De Lima, O. M. Del Cima, E. S. Miranda, {\it On the electron–polaron–electron–polaron scattering and Landau levels in pristine graphene-like quantum electrodynamics}, Eur. Phys. J. B 93, 187 (2020).

    \bibitem{Well21}
    W. B. De Lima, O. M. Del Cima, E. S. Miranda, {\it On the ultraviolet finiteness of parity preserving $U(1) \times U(1)$ massive QED$_3$}, Annals Phys. 430, 168504 (2021).

    \bibitem{Well22}
    W. B. De Lima and P. De Fabritiis, {\it Self-dual Maxwell-Chern-Simons solitons in a parity-invariant scenario}, Phys. Lett. B 833, 137326 (2022).

    \bibitem{Well23}
    W. B. De Lima and P. De Fabritiis, {\it Vortices in a parity-invariant Maxwell–Chern–Simons model}, Mod. Phys. Lett. 38, 2350174 (2023).

%%%%%%%%%%%%%%%%%%%%%%%%%%%%%%%%%%%%%%%%%%%%%%%%%%%%%
    \bibitem{Collins84}
    J. C. Collins, {\it Renormalization: An Introduction to Renormalization, the Renormalization Group and the Operator-Product Expansion}, Cambridge University Press (1984).

    \bibitem{Dudal26c}
    D. Dudal, L. Rosa, and S. Stouten, {\it Examples of Gribov copies in the presence of planar boundaries, and an emergent boundary gauge invariance in Yang-Mills theory}, arXiv:2608.30447.

    \bibitem{Wolfram83}
    J. Ambjørn and S. Wolfram, {\it Properties of the vacuum. I. Mechanical and thermodynamic}, Ann. Phys. 147, 1 (1983).

    \bibitem{vanDam70}
    H. van Dam and M. J. G. Veltman, {\it Massive and massless Yang-Mills and gravitational fields}, Nucl. Phys. B 22, 397 (1970).

    \bibitem{Zakharov70}
    V. I. Zakharov, {\it Linearized gravitation theory and the graviton mass}, JETP Lett. 12, 312 (1970).


%%%%%%%%%%%%%%%%%%%%%%%%%%%%%%%%%%%%%%%%%%%%%%%%%%%%
   
%%%%%%%%%%%%%%%%%%%%%%%%%%%%%%%%%%%%%%%%%%%%%%%%%%%%%
   


   
   
\end{thebibliography}
\end{document}